\documentclass{ws-procs961x669}            

\begin{document}
\title{Tidal Deformability as a Probe of Dark Matter in Neutron Stars}

\author{D. Rafiei Karkevandi$^*$ }

\address{ICRANet-Isfahan, Isfahan University of Technology,\\Isfahan, 84156-83111, Iran\\
$^*$E-mail: davood.rafiei64@gmail.com}

\author{S. Shakeri$^\dagger$}

\address{Department of Physics, Isfahan University of Technology,\\ Isfahan, 84156-83111, Iran,\\
ICRANet-Isfahan, Isfahan University of Technology,\\
$^\dagger$E-mail: s.shakeri@iut.ac.ir}

\author{V. Sagun$^\ddagger$}

\address{CFisUC, Department of Physics, University of Coimbra,\\Coimbra, 3004-516, Portugal\\
$^\ddagger$E-mail: violetta.sagun@uc.pt}

\author{O. Ivanytskyi$^\S$}

\address{Institute of Theoretical Physics, University of Wroclaw,\\Wroclaw, 50-204 , Poland\\
$^\S$E-mail: oleksii.ivanytskyi@uwr.edu.pl}

\begin{abstract}
The  concept of boson stars (BSs) was first introduced by Kaup and Ruffini-Bonazzola in the 1960s. Following  this idea, we  investigate an  effect of self-interacting asymmetric bosonic dark matter (DM) according to  Colpi et al. model for BSs (1986) on different observable properties of neutron stars (NSs). In this paper, the bosonic DM  and  baryonic matter (BM) are mixed together and interact only through gravitational force. The presence of DM as a core of a compact star or as an  extended halo around it is examined by applying different boson masses and DM fractions for a fixed coupling constant.
The impact of DM core/halo formations on a DM admixed NS properties is probed through the maximum mass and tidal deformability of NSs.
Thanks to the recent detection of Gravitational-Waves (GWs) and the latest  X-ray observations, the DM admixed NS's features are compared to LIGO/Virgo and NICER  results.

\end{abstract}


\bodymatter

\section{Introduction}\label{sec1}

The evidence for the existence of dark matter (DM) which constitutes up to $85\%$ of the matter in the universe, is implied from various astrophysical and cosmological observations. However, despite the enormous experimental efforts in the past decades, the nature of these particles remains elusive. In addition to various terrestrial experiments,  compact astrophysical objects such as neutron stars (NSs) can be served as valuable natural  detectors to constrain the properties of DM. The presence of DM in NS interior, depending on the various hypotheses of introducing them and their features, could have signficant effects on the properties of NSs  \cite{Panotopoulos_2017,Nelson_2019,Ellis:2018bkr, PhysRevD.102.063028,Rezaei:2016zje, deLavallaz:2010wp,Gresham:2018rqo,2021arXiv211106197D,Jimenez:2021nmr}. 

There are different scenarios assuming existence of DM  in the NS interior, which are mostly based on DM accretion during different stages of stellar evolution.
The main of these stages are  : a) progenitor, b) main sequence star, c) supernova explosion with formation of a proto-NS, and d) equilibrated NS \cite{Navarro:1995iw,Ruffini_2015,Arg_elles_2018, 2020Univ....6..222D, PhysRevD.102.063028,Ciancarella:2020msu}. 
As an alternative way, DM can be produced during the supernova explosion or NS merger leading to presence of DM in the NSs \cite{Nelson_2019,Ellis:2018bkr}. High level of DM fraction inside NS is reachable through 
other mechanisms such as  (i) dark compact objects or DM clumps as the accretion center of baryonic matter (BM)
 \cite{Ellis:2018bkr,Goldman_2013,Ciarcelluti_2011},
(ii) DM captured by NS from a Dark star companion or even within Dark star–NS merger 
 \cite{Ciarcelluti_2011,Gresham:2018rqo,Sandin_2009,Ellis:2018bkr,Kouvaris:2015rea}, (iii) NS can pass through a  region in the Galaxy with extremely high DM density  leading to accumulation of vast amount of DM \cite{Sandin_2009,DelPopolo:2020pzh,Deliyergiyev_2019,Li:2012ii,DelPopolo:2019nng}.  
 
In order to examine the impact of DM on the NS properties, two main and qualitatively different pictures are considered, a) Self-annihilating DM  affecting luminosity, effective temperature and cooling process of NSs \cite{Kouvaris:2007ay, Bhat:2019tnz,Fuller:2014rza,Acevedo:2020avd,Sedrakian:2015krq,Sedrakian:2018kdm} and b) Asymmetric DM (ADM)  with negligible annihilation rate caused by particle-antiparticle  asymmetry in  the  dark sector \cite{Kaplan:2009ag,Shelton:2010ta,Petraki:2013wwa,Kouvaris:2015rea}. We consider the second possibility allowing stable and massive DM particles to reside in a core of the NS. It was pointed out that the presence of DM particles in the core can significantly decrease the mass of a compact object \cite{Leung:2011zz,Deliyergiyev_2019,PhysRevC.89.025803, Li:2012ii}. However, it was shown that light DM particles form an extended halo around the NS and can increase its gravitational mass~\cite{PhysRevD.102.063028,Nelson_2019}. It is worth mentioning  that both of the aforementioned cases for combination of DM and BM within NS are known as DM admixed NS.

Regarding the ADM model, generally two methods  have been utilized so far to extract the properties of the DM admixed NS from the Tolman-Oppenheimer-Volkof (TOV)
equations \cite{Tolman:1939jz,Oppenheimer:1939ne}. 1) Single fluid formalism, for which an Equation of State (EoS) is considered for the whole star by inserting DM-BM interactions \cite{Panotopoulos_2017,Gresham:2018rqo,Das:2018frc,Das:2020vng,Sen:2021wev,Das:2021wku}. 2) Two-fluid formalism, for which DM and BM interact only through gravitational force, and two individual EoSs have to be considered for the DM and BM fluids \cite{Sandin_2009,Mukhopadhyay:2016dsg,Ciarcelluti_2011,Li_2012,Li1:2012sg,Tolos_2015}.

In this research, we apply a two-fluid formalism for the DM admixed NS. Bosonic DM is described by a complex scalar field with repulsive self-interaction. Historically, this model has been applied to describe  hypothetical self-gravitating objects composed of bosons, so called boson stars. The idea of BS was first proposed by Kaup \cite{Kaup:1968zz} and Ruffini-Bonazzola \cite{Ruffini:1969qy} for non-interacting bosons. The Heisenberg uncertainty principle was the only source of pressure of the BS matter resisting gravitational contraction. This leads to much lower maximum mass of BS compared to the Chandrasekhar mass. The pressure of the BS matter was significantly increased by introducing repulsive self-interaction proposed by Colpi et al  \cite{Colpi:1986ye}. Within this approach, stellar mass objects are supported by the DM particle mass about hundreds MeV and dimensionless coupling constant is of order of unity (see a comprehensive review on BSs in \cite{Schunck:2003kk,Liebling:2012fv,Visinelli:2021uve}). Another component, i.e. BM, is modeled by the induced surface tension (IST) EoS. It was successfully applied to describe the nuclear matter, heavy-ion collision data and dense matter existing inside NS \cite{2018NuPhA.970..133B, Sagun:2018cpi, Sagun11:2018sps}.

With this set up we study the effects of self-repulsive bosonic ADM on the mass-radius (M-R) profile and tidal deformability parameter \cite{Nelson_2019, Ellis:2018bkr, Quddus:2019ghy, Ciancarella:2020msu, Zhang:2020dfi, LeTiec:2020spy, Das_2019, Sen:2021wev,Das:2021wku} inferred from the GW signals related to post-merger stages of NSs \cite{Ellis:2017jgp, Bezares:2019jcb, Bezares:2018qwa, Horowitz:2019aim, Bauswein:2020kor}. Such a combined analysis based on the recent LIGO/Virgo results \cite{LIGOScientific:2017vwq,LIGOScientific:2020aai,LIGOScientific:2017vwq,LIGOScientific:2020aai} opens a new possibility to study the internal structure of compact objects which may contain DM. 
 
To be specific, in this work we consider a model of Colpi et al. \cite{Colpi:1986ye} with sub-GeV DM particles of mass $m_{\chi}\sim\mathcal{O}$(100 MeV) and self-coupling constant $\lambda=\pi$. We analyze two key observational constraints of NSs, i.e. maximum mass and tidal deformability. The first of them is based on the NICER observation of the heaviest known pulsar PSR J0740+6620 with mass  $2.072^{+0.067}_{-0.066}M_{\odot}$   \cite{Riley:2021pdl} and corresponds to requiring the maximum stellar mass to be at least $M_{max}=2M_\odot$. The merger event GW170817\cite{Abbott_2017} leads to the second constraint on the  dimensionless tidal deformability  $\Lambda\leq580$  for $M=1.4M_\odot$ \cite{Abbott:2018exr}. Using these constraints we probe DM admixed NSs at various masses and fractions of DM.

The rest of the paper  is organized as follows. In Sec. \ref{sec2} the DM and BM EoSs are described. In Sec. \ref{sec3} we explain the two-fluid TOV formalism, DM halo and DM core formations and their impacts on the M-R profile of DM admixed NSs. Sec. \ref{sec4} is devoted to probing the effect of DM halo/core configurations on the tidal deformability. Our
conclusions will be presented in Sec.\ref{sec5}. We use units in which $\hbar=c=G=1$.

\section{Bosonic DM and BM models}\label{sec2}

\subsection{Dark Boson Star}

In  the  following  we  apply  a  model  of complex  scalar field as the bosonic DM with repulsive self-interaction potential, $V(\phi)=\frac{\lambda}{4}|\phi|^4$,  minimally coupled to gravity and described by the action

\begin{eqnarray}\label{action}
S=\int d^{4}x \sqrt{-g}
\Bigg[\frac{M_{Pl}^{2}}{2}R - \frac{1}{2}\partial_{\mu}\phi \partial^{\mu}\phi^{\ast}
-\frac{1}{2}m_{\chi}^{2}|\phi|^{2} -\frac{1}{4} \lambda |\phi|^{4} \Bigg]\,,
\end{eqnarray}
    where $m_{\chi}$ is the boson mass, $\lambda$ stands for the dimensionless coupling constant and $M_{Pl}$ corresponds to the Planck mass  \cite{Colpi:1986ye,Maselli:2017vfi}. In this setup a coherent scalar field  is governed by  both Klein-Gordon and Einstein equations which can potentially form Bose-Einstein Condensate (BEC) if the temperature is sufficiently low \cite{PhysRevD.68.023511,Suarez:2013iw}. It has been assumed a spherically symmetric configuration for the scalar field $\phi(r,t)=\Phi (r) e^{i\omega t}$  and a static metric to rewrite  Klein-Gordon-Einstein (K.G.E)  equations  to a set of ordinary differential equations \cite{Colpi:1986ye}. In a parameter region for $\lambda$ as
\begin{eqnarray}
\lambda \gg 4\pi (m_{\chi}/M_{Pl})^{2}=8.43\times 10^{-36} \left( \frac{m_{\chi}}{100\,\text{MeV}} \right)^{2} \,
\end{eqnarray}
the scalar field only varies on a large length scale $R\sim (\frac{\lambda M_{Pl}^{2}}{4\pi m_{\chi}^{4} })^{1/2}$ and we can ignore spatial derivatives in the scalar field equation and solve it locally. In this limit which is called strong coupling regime, the system can be approximated as a prefect fluid and the anisotropy of pressure will be ignored \cite{Mielke:2000mh,Schunck:2003kk,Chavanis:2011cz}. This leads to the following EoS describing a self-interacting and self-gravitating bosonic system so-called BS
\begin{equation}\label{e1}
P=\frac{m_{\chi}^{4}}{9\lambda}\left( \sqrt{1+\frac{3\lambda}{m_{\chi}^{4}}\rho}-1\right)^{2}.
\end{equation}
We recently presented an alternative derivation of this EoS in locally flat space-time by using the mean-field approximation (see appendix of  \cite{Karkevandi:2021ygv}). Stellar mass BSs can be formed for $\lambda\sim\mathcal{O}(1)$ and $m_{\chi}\sim\mathcal{O}$(100 MeV)   \cite{Schunck:2003kk,Pacilio:2020jza}, in this section, we focus on this range of model parameters.

\begin{figure}[h]%
\begin{center}
  \parbox{2.4in}{\includegraphics[width=2.4in]{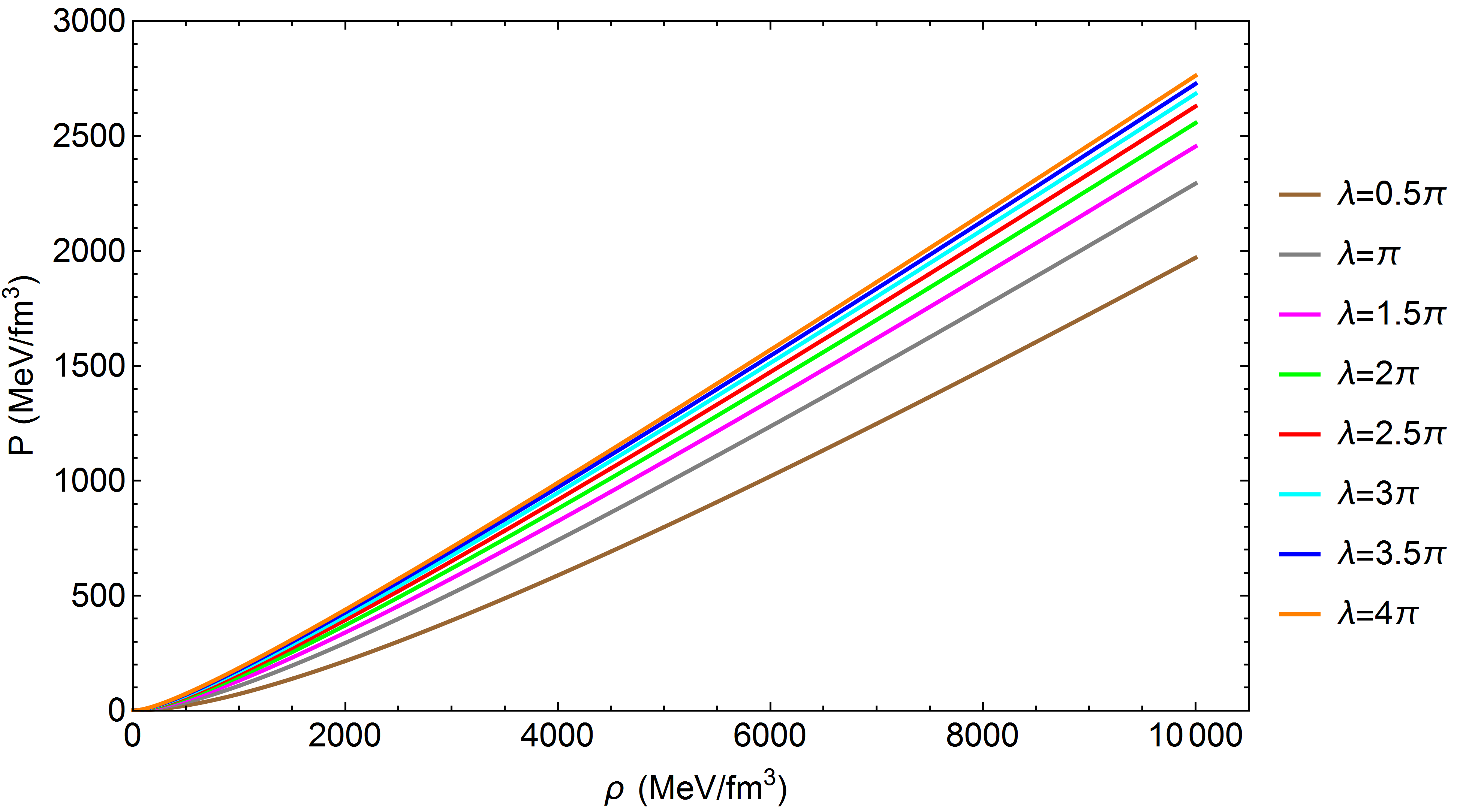}}
  \hspace*{4pt}
  \parbox{2.4in}{\includegraphics[width=2.4in]{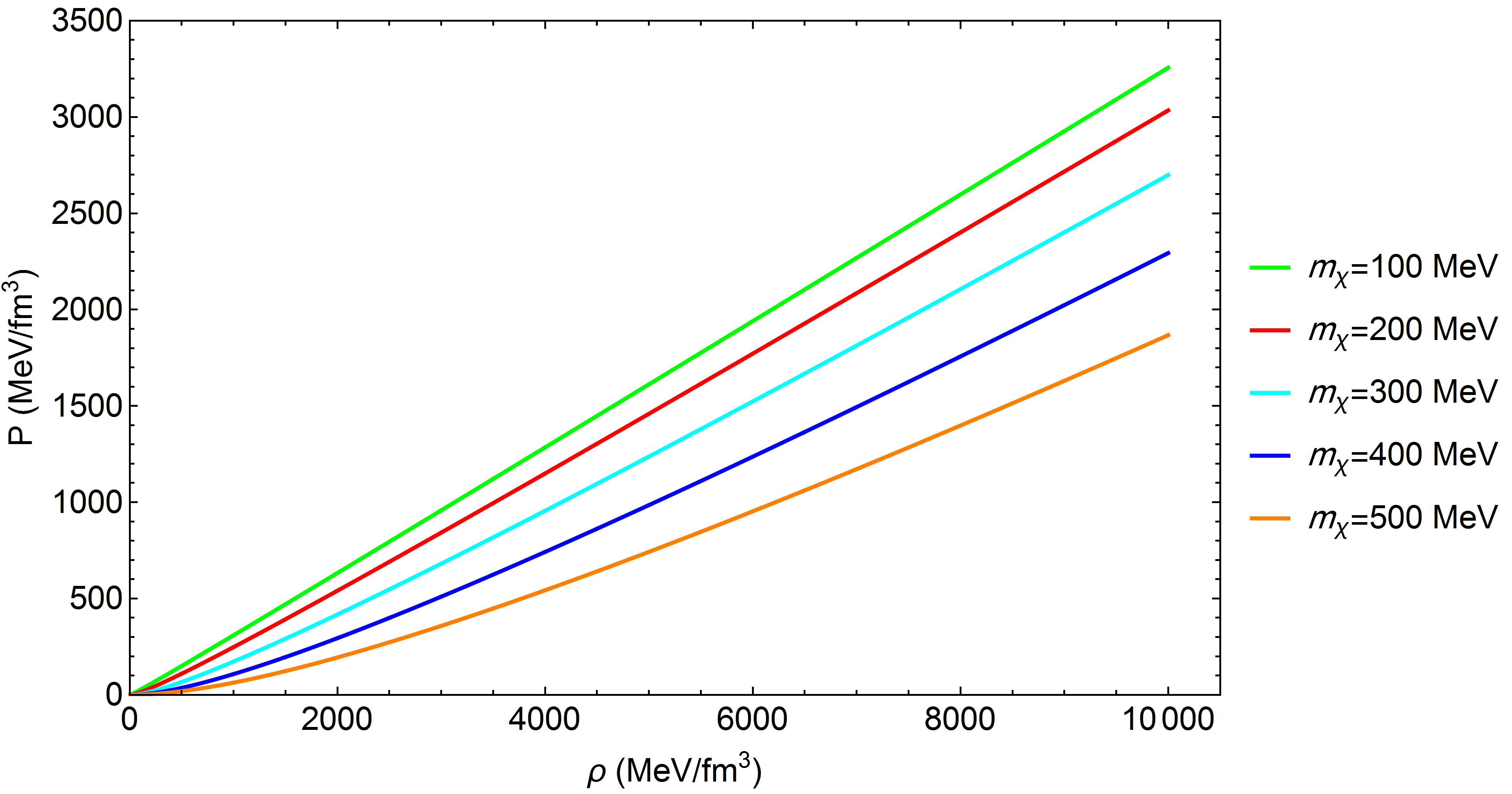}}
  \caption{Pressure as a function of density for bosonic matter obtained for $\lambda=\pi$ and  various DM masses as labeled (right); for $m_{\chi}=400$ MeV and different values of coupling constant (left).}%
  \label{fig1}
\end{center}
\end{figure}
Fig. \ref{fig1} shows pressure of bosonic matter as a function of density for different masses and coupling constants as labeled. As it is shown in Fig. \ref{fig1}, (right panel) the pressure decreases with increasing of  boson masses and (left panel) the pressure rises with the enhancement of the coupling constant or equivalently increasing the repulsive force between bosons.

In different density limits one can approximate Eq. (\ref{e1}) in a typical polytrope form $P=K\rho^{\gamma}$, where polytropic index at low density $\gamma\simeq 2$  and it smoothly reaches to $\gamma\simeq 1$ at high density. 
At low density regime, the bosonic DM EoS, Eq. (\ref{e1}), is reduced to 
\begin{eqnarray}\label{e2}
P&\approx& \frac{\lambda}{4m_{\chi}^4}\rho^{2}\, .
\end{eqnarray}

However,  for high density regime or correspondingly for very light bosons or high coupling constant, Eq. (\ref{e1}) reaches to radiation EoS with $P\approx \rho/3 $. Similar equation to Eq. (\ref{e2}) has been obtained so far for a dilute  self-interacting boson gas in a  self-gravitating system (BS) known as Gross-Pitaevskii-Poisson (G.P.P)
equation \cite{Chavanis:2011cz,Li:2012sg,Chavanis:2019bnu}. In fact, the G.P.P equation describes the BEC phase in a dilute gas where only two-body mean field interaction is considered near zero temperature \cite{Dalfovo_1999,Rogel_Salazar_2013,pethick_smith_2008}.

\begin{figure}[h]%
\begin{center}
  \parbox{2.4in}{\includegraphics[width=2.4in]{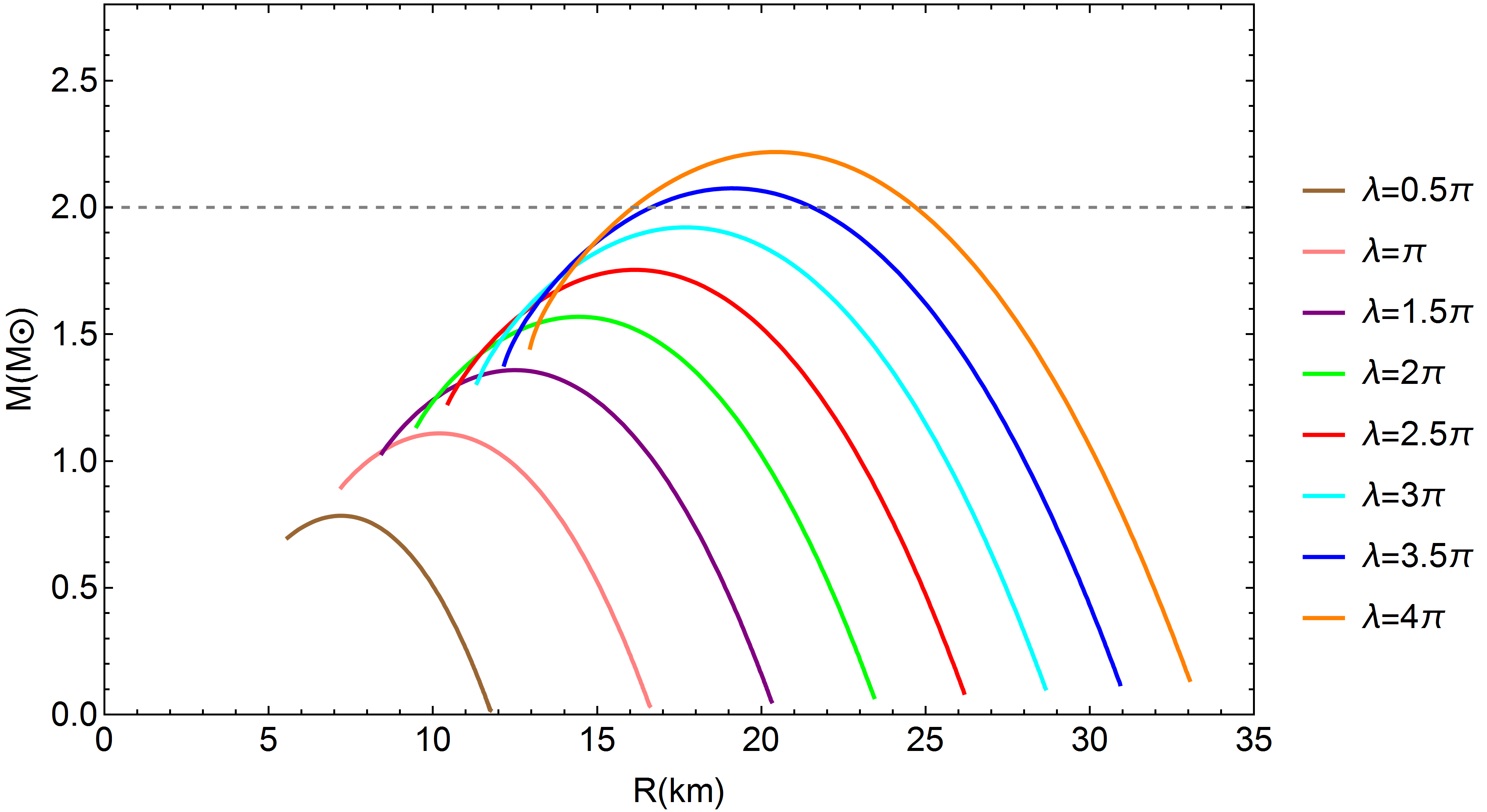}}
  \hspace*{4pt}
  \parbox{2.4in}{\includegraphics[width=2.4in]{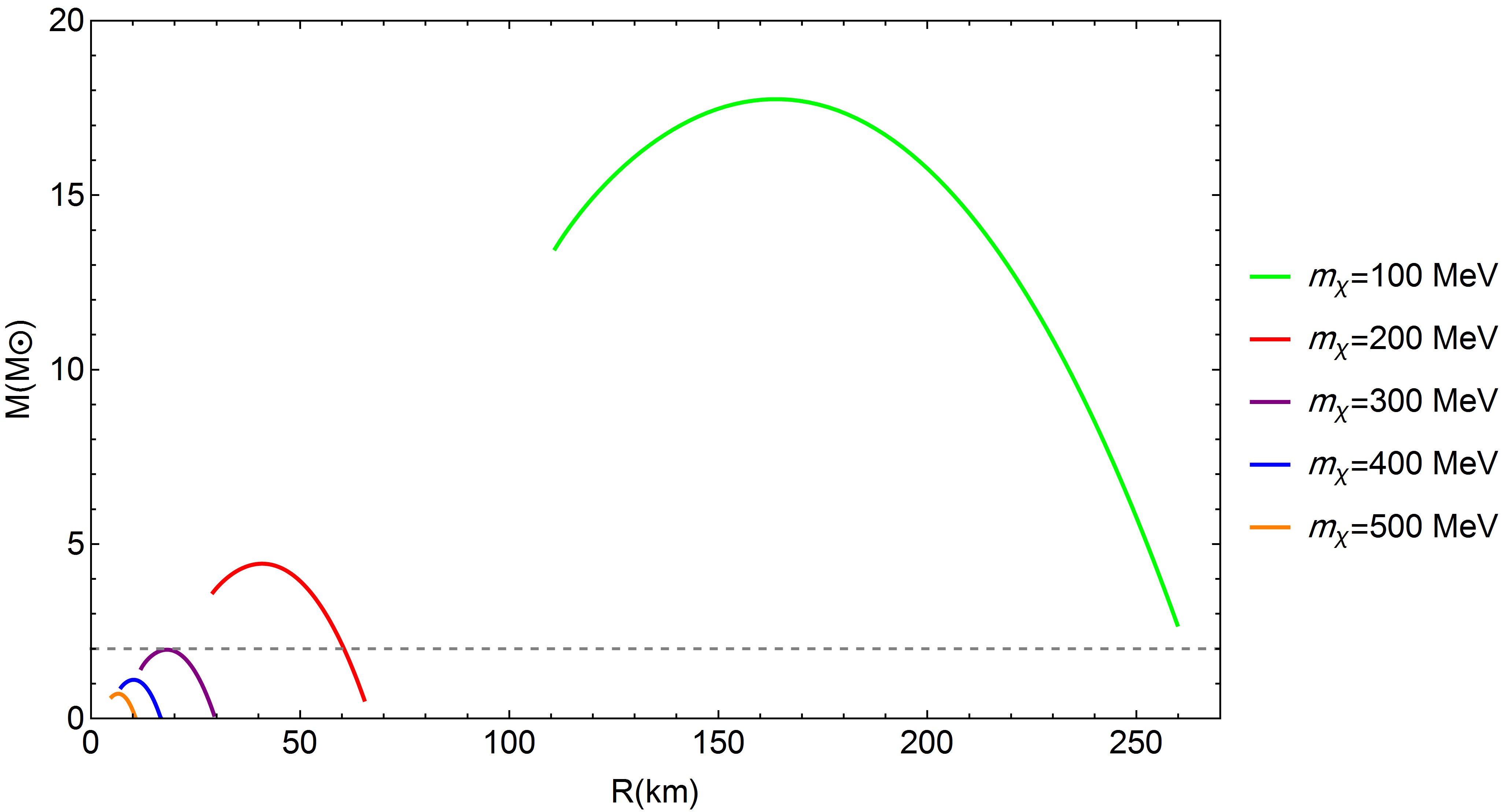}}
  \caption{M-R profile for BSs based on Eq. (\ref{e1}), (right panel) for the fixed value of coupling constant $\lambda=\pi$ and different boson masses. (Left panel) Calculations are made for fixed boson mass $m_{\chi}=400$ MeV and different values of coupling constant as labeled at the figure. The gray dashed line shows the $2M_\odot$ limit.}%
  \label{fig3}
\end{center}
\end{figure}

In   Fig. \ref{fig3}, we present the M-R diagrams of BSs obtained by solving TOV  equation for Eq.  (\ref{e1}). As it is indicated in the right panel, by decreasing the boson mass, the maximum gravitational mass of BSs increases and even goes above $2M_{\odot}$ \cite{Antoniadis:2013pzd,Cromartie:2019kug,Riley:2021pdl} and the corresponding radius goes well above typical NS radius.
In the left panel, it is shown that higher self-coupling constant at fixed mass $m_{\chi}=400$ MeV leads to higher maximum masses of BSs. Both the  decreasing of boson mass and increasing of the coupling constant cause an enhancement  in pressure of the system and consequently the rise of the maximum mass and radius of BSs.

\begin{figure}[h]%
\begin{center}
  \parbox{2.4in}{\includegraphics[width=2.4in]{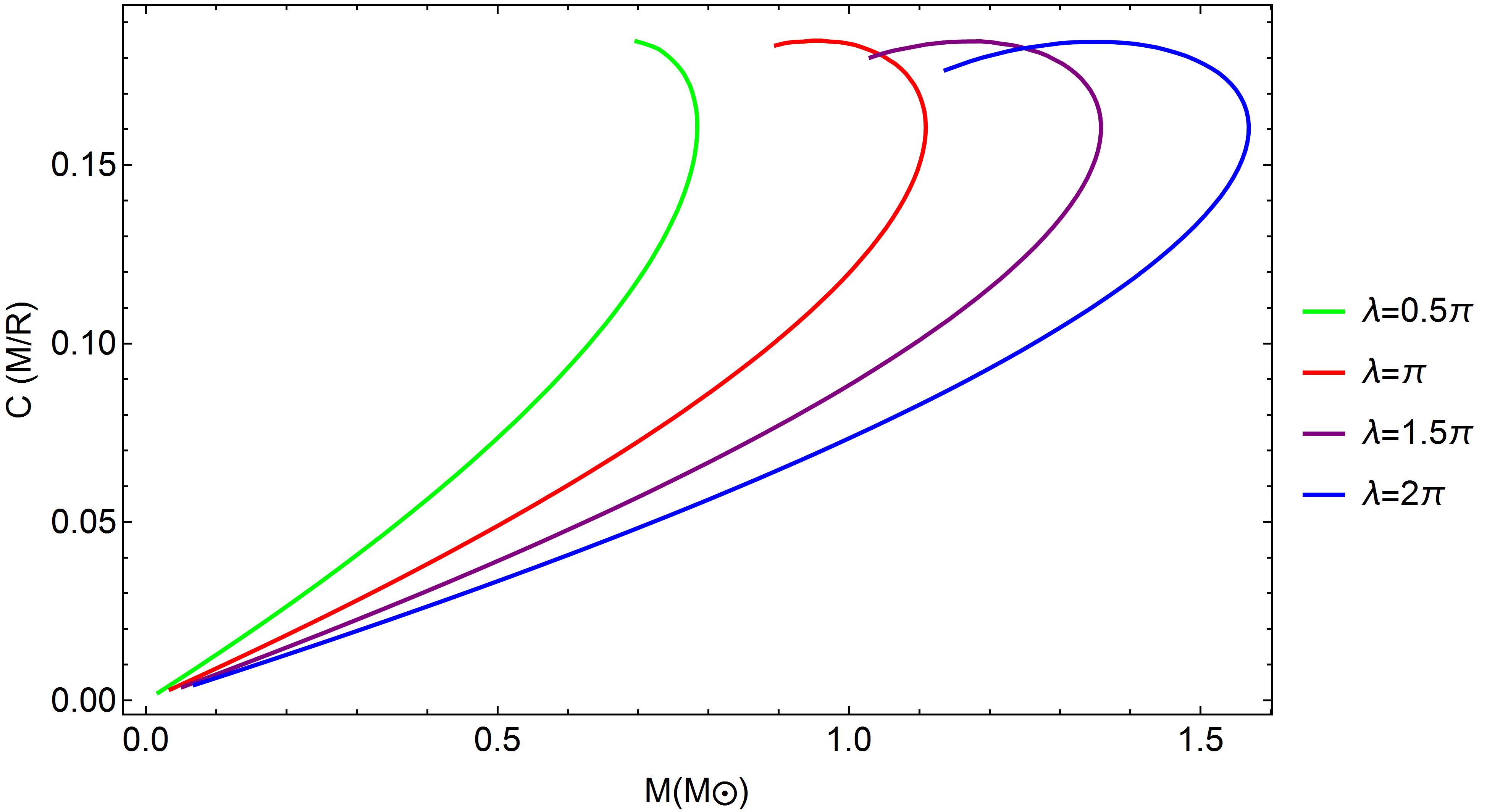}}
  \hspace*{4pt}
  \parbox{2.4in}{\includegraphics[width=2.4in]{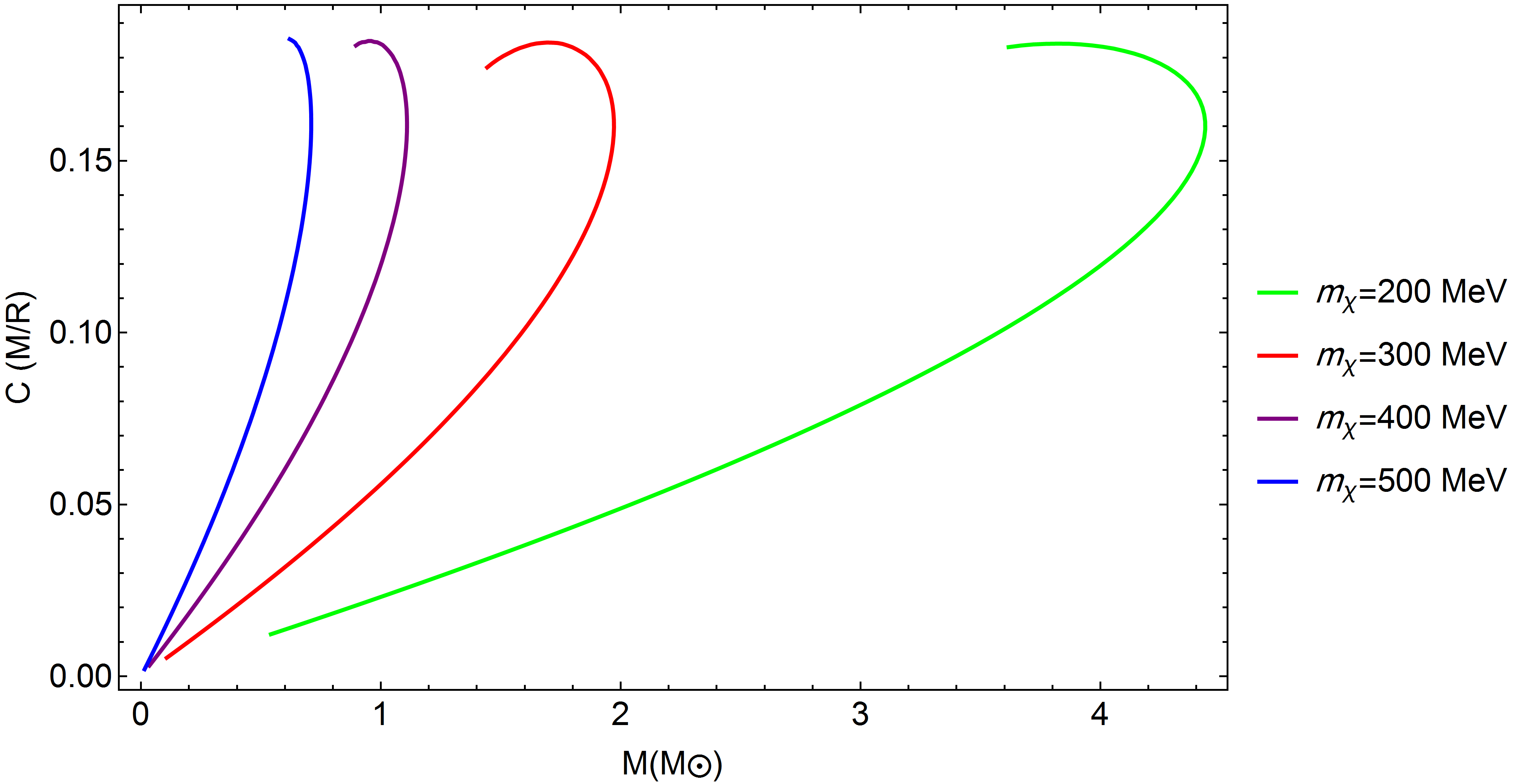}}
  \caption{Compactness of BSs as a function of their mass obtained for a fixed coupling constant $\lambda=\pi$ and different values of boson mass (right); fixed particle's mass $m_{\chi}=400 $ MeV and various $\lambda$ values (left).}
  \label{fig4}
\end{center}
\end{figure}

Moreover, the variation of compactness $\mathcal{C}=M/R$ with respect to mass of BSs for different values of $m_{\chi}$ and  $\lambda$ is presented in Fig. \ref{fig4}. It shows the same dimensionless maximum compactness $\mathcal{C}_{(max)}\simeq0.16$ for all cases. We see that the maximum compactness of a BS based on Eq. (\ref{e1}) is independent of free parameters of the model, namely $m_{\chi}$ and $\lambda$ \cite{} and for all the parameter space is well below the black hole formation limit $\mathcal{C}=0.5$ \ \  \cite{Liebling:2012fv,AmaroSeoane:2010qx}.

\subsection{Neutron Star}

\begin{figure}[h]%
\begin{center}
  \parbox{2.4in}{\includegraphics[width=2.4in]{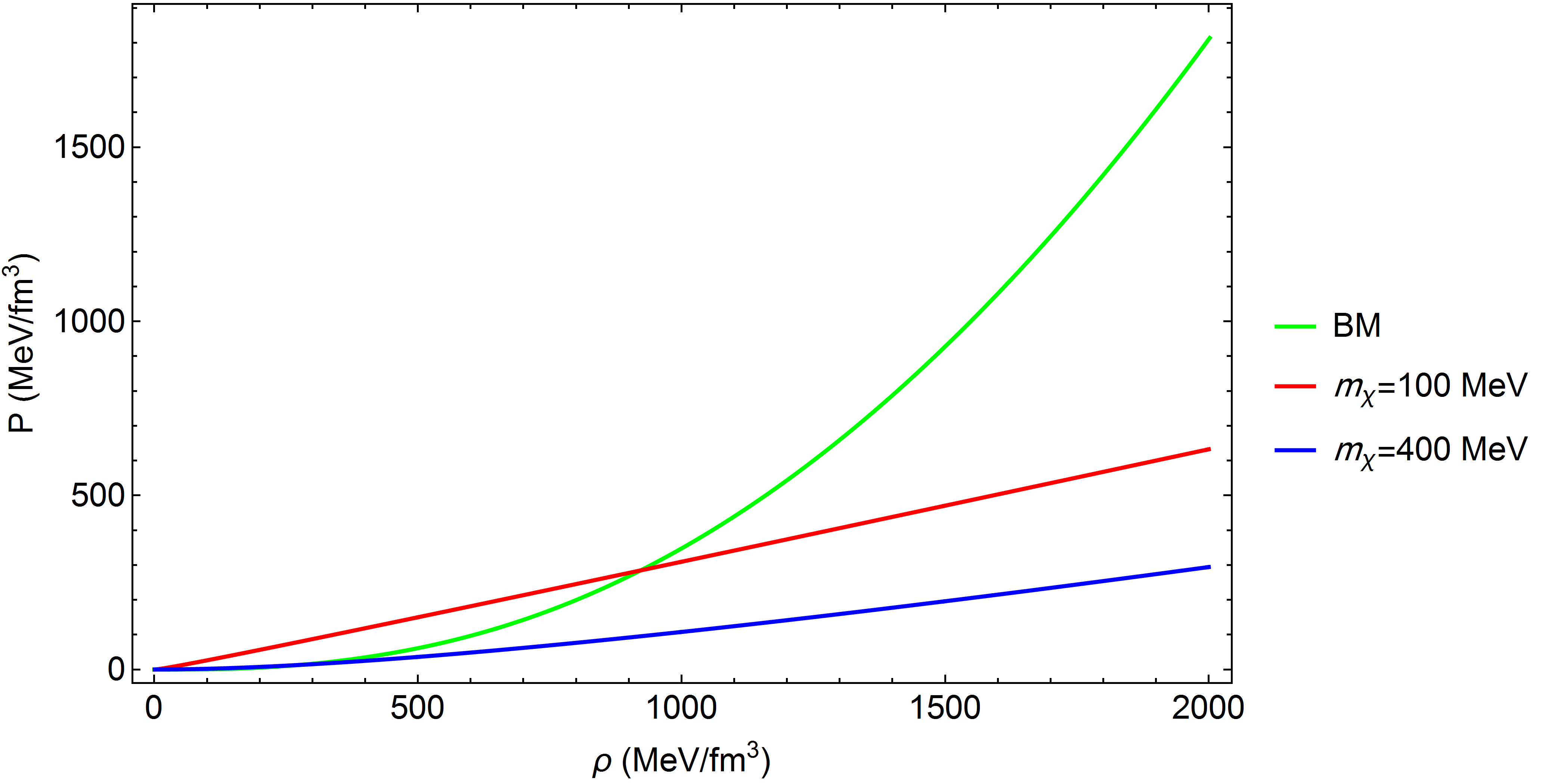}}
  \hspace*{4pt}
  \parbox{2.4in}{\includegraphics[width=2.4in]{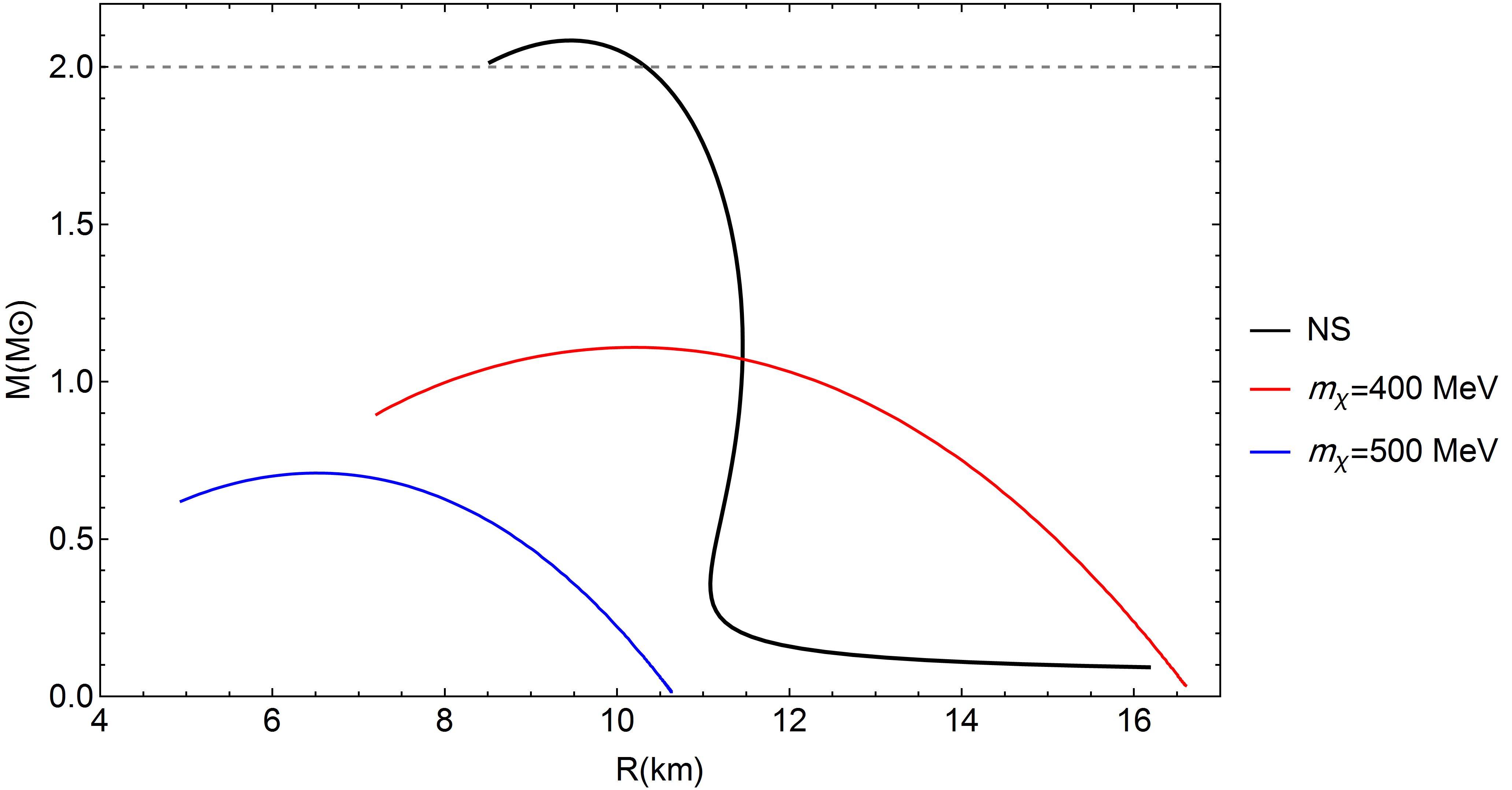}}
  \caption{Comparing BM and DM EoSs, left panel shows pressure vs. energy density for two different values of boson mass and $\lambda=\pi$. Right panel indicates M-R profiles for the NSs and BSs, considering $m_{\chi}=400,500$ MeV and the same coupling constant.}%
  \label{fig2}
\end{center}
\end{figure}

For the baryon component (NS matter), we use the unified EoS with induced surface tension (IST)  where both the short-range repulsion and long-range attraction between baryons have been taken into account \cite{Sagun:2018cpi,Sagun11:2018sps}. The IST EoS reproduces the nuclear matter properties \cite{2014NuPhA.924...24S}, fulfills the proton flow constraint \cite{Ivanytskyi:2017pkt}, provides a high-quality description of hadron multiplicities created during the nuclear-nuclear collision experiments \cite{2018EPJA...54..100S} as well as the matter inside compact stars \cite{2018NuPhA.970..133B, Sagun:2018cpi, Sagun11:2018sps}. The EoS is in a very good agreement with the latest NS observations providing the maximum mass $M_{max}=2.08 M_{\odot}$ and radius of the $1.4 M_{\odot}$ star equals to $R_{1.4}=11.37$ km \cite{Sagun:2020qvc}. In our work the crust part of the NS's EoS is described via the polytropic EoS with $\gamma=4/3$ \ \ \cite{PhysRevD.102.063028}.  
In Fig. \ref{fig2}, the change of pressure for a same density regime is plotted for BM and DM EoSs (left panel) and we show the M-R profiles of the NS and BS based on our considered EoSs (right panel).

\section{Two-fluid TOV equations and Maximum mass}\label{sec3}

\begin{figure}[h]
\begin{center}
\includegraphics[width=4in]{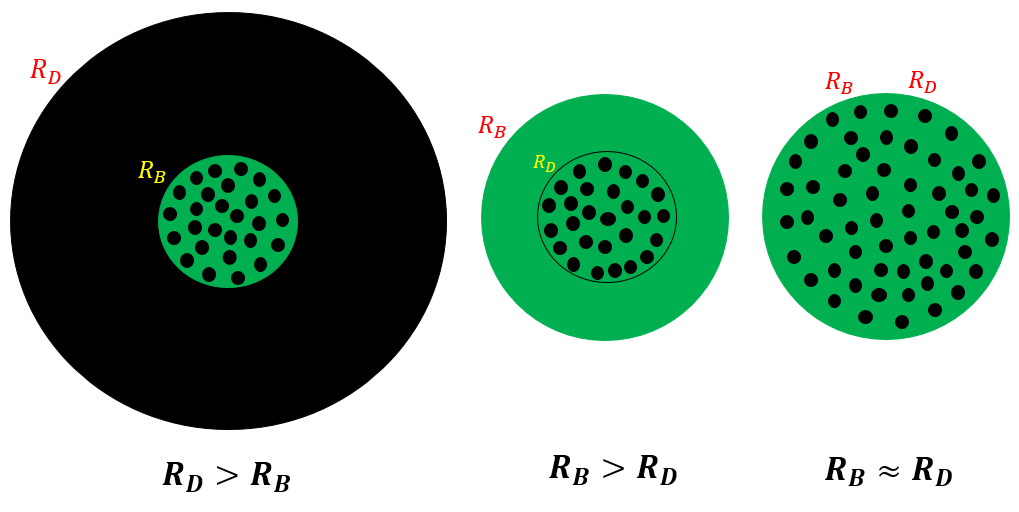}
\end{center}
\caption{Three possible configurations of a DM admixed NS, (left) DM halo, (middle) DM core and (right) DM  
is distributed in a whole NS. Note that for the DM core and halo cases the core of the object is a mixture of BM and DM. Green and black colors denote BM and DM, respectively. }
\label{configu}
\end{figure}

In order to study compact objects formed by the admixture of BM and DM that
interact only through gravity, we use two-fluid TOV formalism  \cite{Sandin_2009,Ciarcelluti_2011} shown by  

\begin{eqnarray}\label{e6}
\frac{dp_{\text{B}}}{dr}&=& -\left( p_{\text{B}} +\epsilon_{\text{B}} \right) \frac{M+4\pi r^{3}p}{r(r-2M)}\,,\\ 
\frac{dp_{\text{D}}}{dr}&=& -\left( p_{\text{D}} +\epsilon_{\text{D}} \right) \frac{M+4\pi r^{3}p}{r(r-2M)}, \label{e6d}
\end{eqnarray}
Here $p=p_{B}+p_{D}$ and $M=M_{T}=\int_{0}^{r} 4\pi r^{2} \epsilon_B (r) dr + \int_{0}^{r} 4\pi r^{2} \epsilon_D (r) dr.$ It can be seen that the total pressure and mass of the object have two contributions from BM and DM fluids shown by B and D indices. In order to solve two-fluid TOV equations two central conditions related to both of the fluids have to be considered. By fixing two central pressures ($p_{B}$ and $p_{D}$) together with the initial conditions at the center  of the star ($M_{B}(r\simeq0) = M_{D}(r\simeq0) \simeq 0$) the  Eqs. (\ref{e6}-\ref{e6d}) are numerically integrated up to the radius at which the  pressure of one of the components vanishes. In principle this radius can be realized as DM radius $R_{D}$ or BM radius $R_{B}$. In the former case the DM  distributed only inside the core while BM extends to larger radius ($R_{B}>R_{D}$), then we set $p_{D}(r>R_{D})=0$ and continue the numerical integration  
to reach the visible radius of the star where  $p_B(R_B)=0$. When we have a BM core, DM can exist as an extended halo around the core with $R_{D}>R_{B}$, where $p_{B}(r>R_{B})=0$. It should be mentioned that for both DM core and DM halo cases, the core of the object is a mixture of DM and BM. Based on our extensive analysis there is another possibility of DM admixed NSs' configurations for which $R_{B}\approx R_{D}$ and DM distributed within the entire NS (see Fig. \ref{configu}).

For all of the possible DM admixed NS configurations, the total gravitational mass of the mixed object is
\begin{eqnarray}
M_{T}=M_{B} (R_{B})+M_{D}(R_{D}).
\end{eqnarray}
However, the observable radius of the star is still defined by $R_{B}$, this is due to the visibility of $R_{B}$ compare to $R_D$ and technical difficulties of indirect detection of dark radius $R_{D}$.  Furthermore, the DM fraction that determines the amount of DM in a DM admixed NS is defined as
\begin{equation}
    F_{\chi}=\frac{M_D(R_{D})}{M_T}.
\end{equation}

\begin{figure}[h]%
\begin{center}
  \parbox{2.4in}{\includegraphics[width=2.4in]{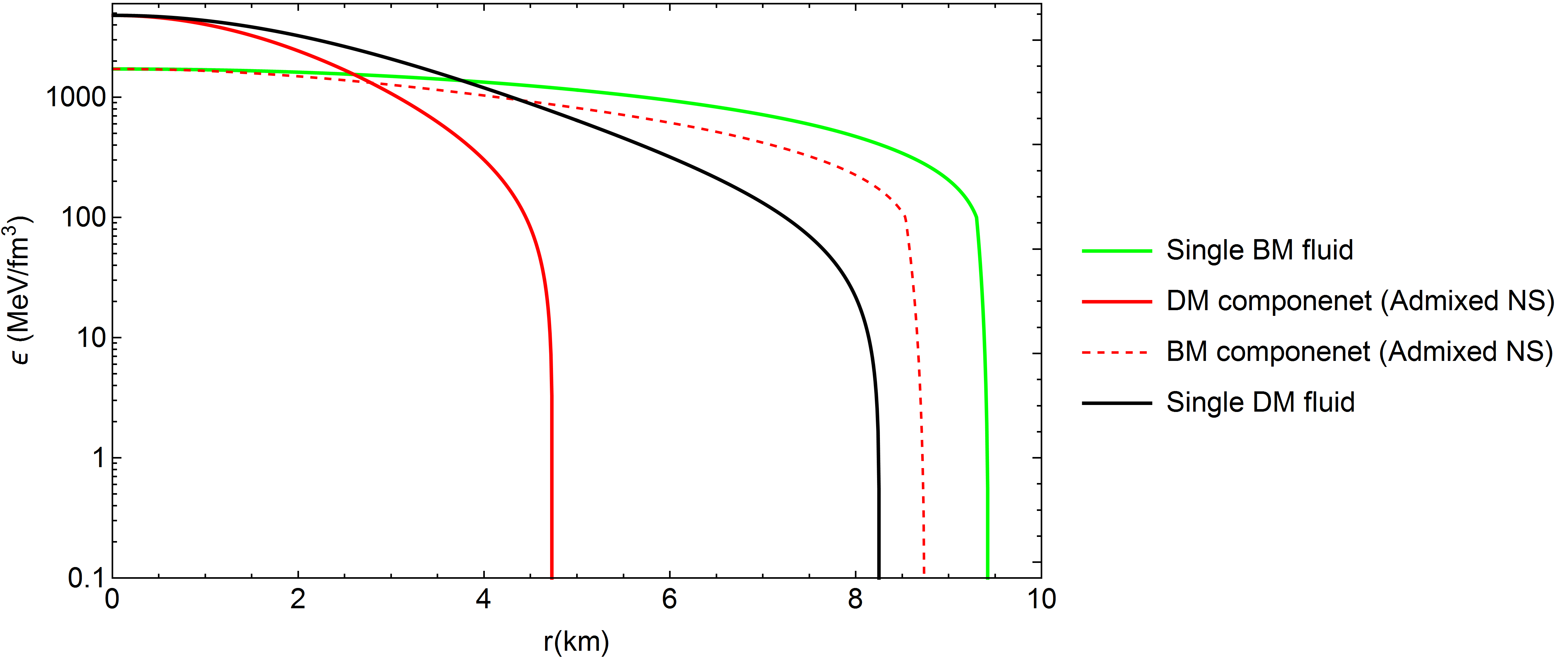}}
  \hspace*{4pt}
  \parbox{2.4in}{\includegraphics[width=2.4in]{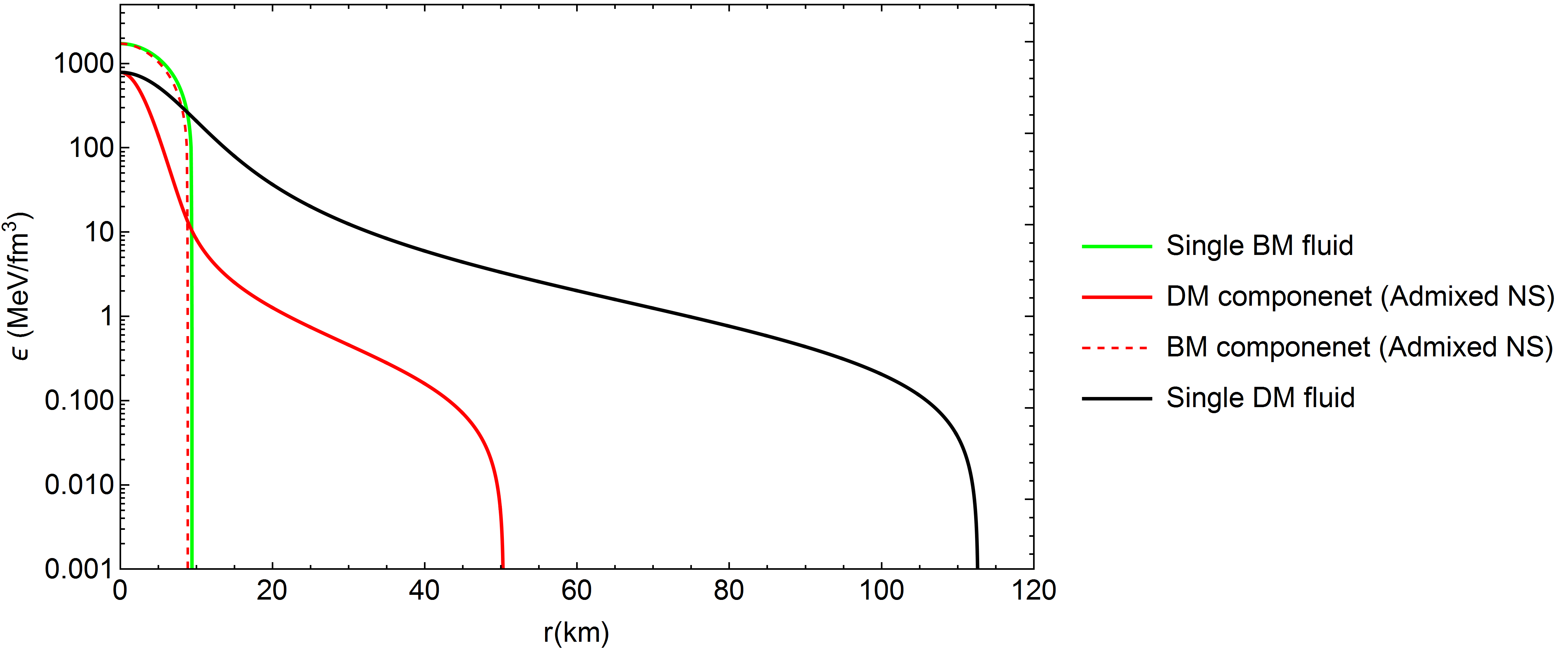}}
  \caption{Energy density  profiles for pure BM and DM stars (green and black curves) shown together with the slitted  DM and BM components of a DM admixed NS (solid and dashed red curves). Left panel corresponds to a DM core formation, while the right one to a DM halo, for $m_{\chi}=400$ MeV and $m_{\chi}=100$ MeV, respectively. For both of the cases, coupling constant is fixed at $\lambda=\pi$ and $F_{\chi}=20\%$.}%
  \label{fig7}
\end{center}
\end{figure}

Hereafter an effect of DM on NS properties is studied for $m_{\chi}$ of about hundreds MeV and
and fixed coupling constant $\lambda=\pi$. Fig. \ref{fig7} shows energy density profiles for a DM admixed NS where the BM (dashed red curves) and DM (solid red curves) components are plotted separately. The energy density profiles for pure BM and DM stars are presented by solid green and black curves, respectively. Here we consider $\lambda=\pi$ and $F_{\chi}=20\%$, the central values of pressure for  BM and DM components are chosen in such a way that   the desired DM fraction $F_{\chi}$ has been obtained.

On the left panel which is obtained for $m_{\chi}=400$ MeV, a DM core with $R_{D}\approx 5~\text{km}$ is embedded in a BM structure with a larger radius. On the right panel, we fixed DM mass to $m_{\chi}=100$ MeV which leads to the formation of a DM halo around the BM fluid with much larger radius. Interestingly, we see that for both DM core and DM halo formations, a reduction occurs in the energy density and the radius of DM and BM fluids in the mixed object compare to pure BM/DM star. This effect is much larger for the DM component and shows that the properties of the single DM fluid have significant effects on the admixed NS and in fact  underlie their features.

By comparing the left and right panels of Fig. \ref{fig7}, we see a transition from DM core to DM halo by changing $m_{\chi}$ from $400$ MeV to $100$ MeV. Therefore, by a  thorough analysis of an effect of model parameters, as a general behaviour, we can conclude that light DM particles with $m_{\chi}<200$ MeV tend to form a halo around a NS, while heavier ones, for low DM fractions, would mainly create a DM core inside a compact star (for more detailed analysis see  \cite{Karkevandi:2021ygv}).

\begin{figure}[h]%
\begin{center}
  \parbox{2.4in}{\includegraphics[width=2.4in]{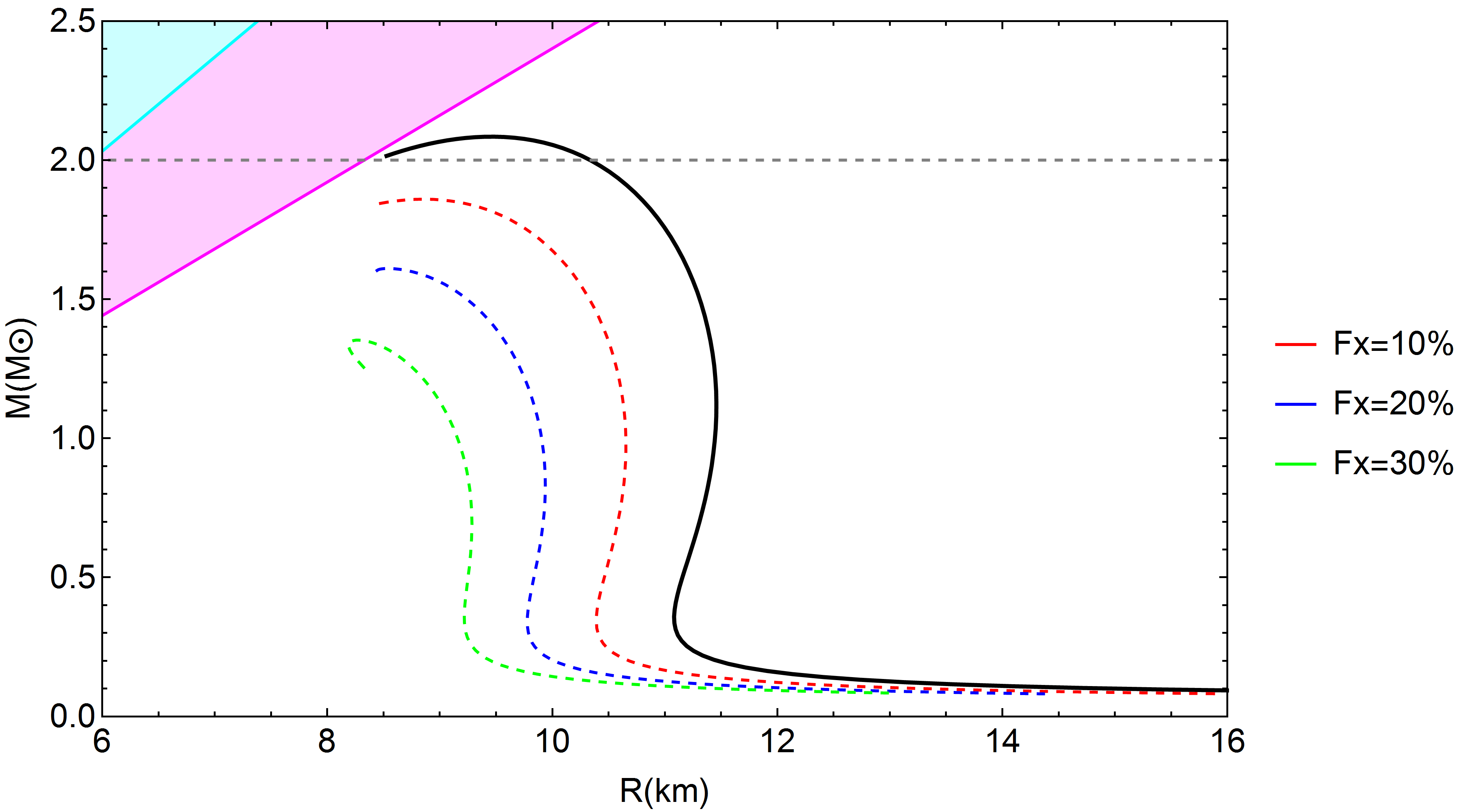}}
  \hspace*{4pt}
  \parbox{2.4in}{\includegraphics[width=2.4in]{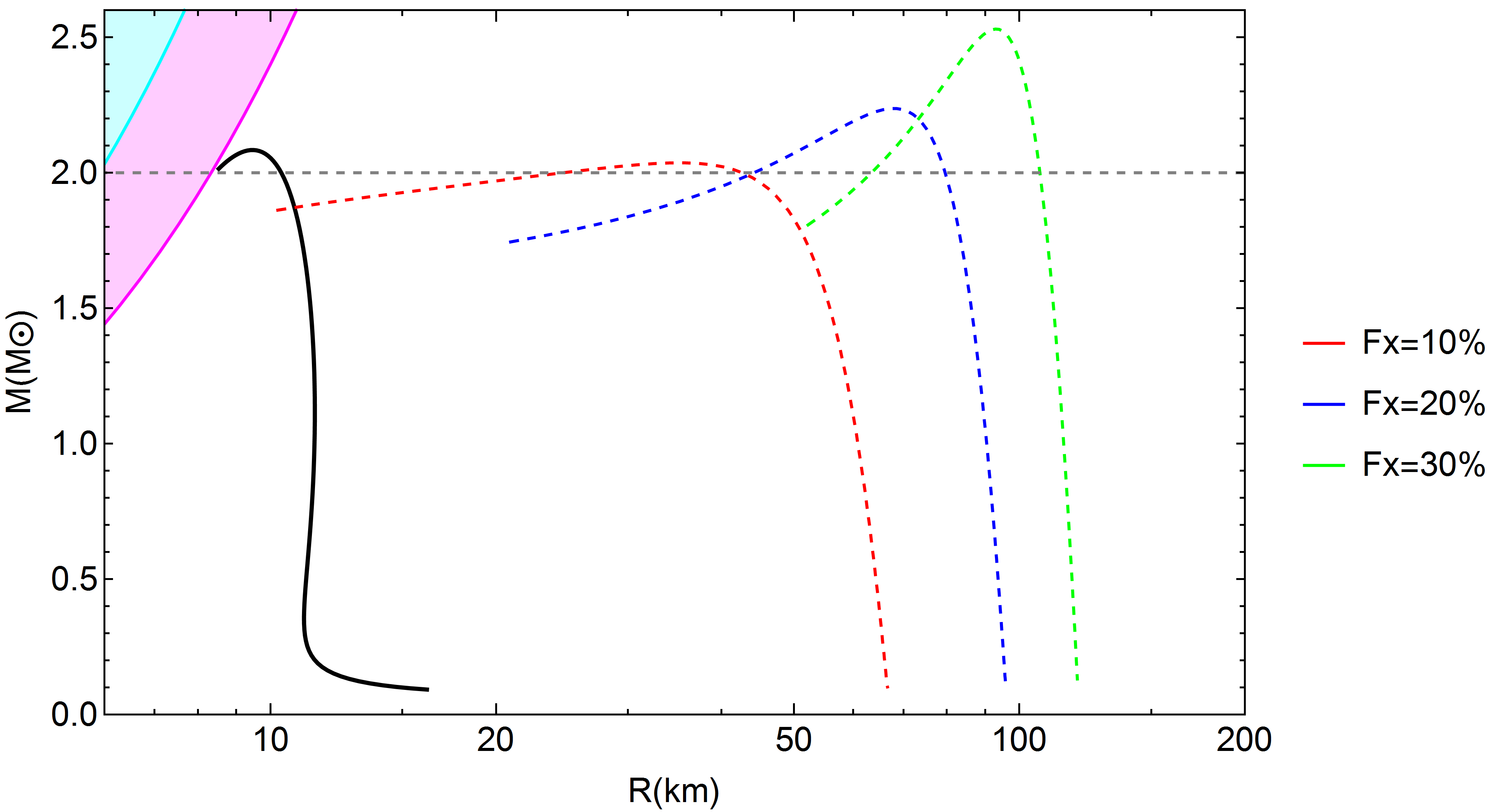}}
  \caption{Mass-Radius profiles for DM admixed NSs for $m_{\chi}=400$ MeV (left) which corresponds to a DM core formation and $m_{\chi}=100$ MeV (right)  that represents an extended DM halo formation around a NS. Coupling constant is fixed to $\lambda=\pi$ and different $F_{\chi}$ are considered as labeled. }%
  \label{fig8}
\end{center}
\end{figure}

The M-R profiles for DM admixed NSs are shown in Fig. \ref{fig8} in which $M=M_{B}+M_{D}$. Here $R$ is  the outermost radius of the star  which is determined by  $R_{B}$ for the DM core  and $R_{D}$ for  the DM halo. The solid black curve shows the M-R relation for the BM fluid (without DM), the gray dashed line indicates the $2M_\odot$ constraint on maximum mass of NSs and the shaded regions colored in magenta and cyan denote the causality and GR limits, respectively. 

As it is shown in Fig. \ref{fig8}, two different boson masses, 100 MeV and 400 MeV lead to a DM halo and a DM core formations, respectively. In the left panel ($m_{\chi}=400$ MeV), it is indicated that DM core formation causes a decrease of the maximum mass well below the $2M_{\odot}$ constraint \cite{Antoniadis:2013pzd,Cromartie:2019kug,Riley:2021pdl}, and also a reduction of the corresponding radius. However, for $m_{\chi}=100$ MeV (see the right panel in Fig. \ref{fig8}) for which a DM halo is formed around a NS, both maximum mass and radius are increased.  Regarding the radius of the object for DM halo formation, it is increased significantly since is determined by $R_{D}$. It is seen that higher DM fractions enhance both of the above behaviours for DM core/halo structures.
Note that for the cases in which a DM halo is formed ($R_D>R_B$), the visible radius of the star remains to be $R_{B}$.

To summarize, we see here that an effect of bosonic DM with repulsive self-interaction onto NSs is in agreement with previous studies which considered different DM models. Thus, an existence of a DM core decreases the maximum  star's mass and the corresponding  radius, while the formation of a DM halo increases these quantities \cite{Leung:2011zz,Nelson_2019,PhysRevD.102.063028,Ellis:2018bkr}. In this regard, the most massive NS observed by NICER \cite{Riley:2021pdl}, PSR J0740+6620 ($M_{max}\simeq2M_\odot$) is compatible with the DM admixed NS scenario.

\section{Tidal deformability of a DM admixed NS}\label{sec4}

GW signal from NS-NS mergers introduce tidal deformability  as a new observable quantity to probe the internal structure of NSs and constrain their macroscopic features  \cite{Raithel:2018ncd,Chatziioannou:2020pqz,Most:2018hfd,De:2018uhw}.
In this section, we analyse an impact of self-interacting bosonic DM on the tidal deformability of a DM admixed NS.

The idea of tidal deformability was first proposed by Tanja Hinderer in 2008 \cite{Hinderer:2007mb,Hinderer:2009ca} which comes from the fact that in a binary system of NSs both of the objects are deformed owing to the imposed tidal forces \cite{Postnikov:2010yn,Zhao:2018nyf,Han:2018mtj}. The tidal deformability expresses the ability of the gravitational field to change the quadrupole structure of a NS which alter the rotational phase of the binary system. Therefore, the  GW signal is influenced  during the inspiral  phase due to the deformation effects of NSs  when  the binary  orbital  radius  becomes  comparable  to  the  radius  of  NS. In fact, taking tidal deformability into account produces a phase shift in GW signal and accelerates the inspiral which leads to an earlier merging\cite{Chatziioannou:2020pqz,Das:2021wku,Lackey:2014fwa}.

The induced quadrupole moment $Q_{ij}$ of a NS due to the external tidal field of its companion $\mathcal{E}_{ij}$  can be parameterized as follows\cite{Hinderer:2007mb,Postnikov:2010yn} 

\begin{eqnarray}\label{2tidal}
Q_{ij}=\lambda_{t} \mathcal{E}_{ij}\, ,
\end{eqnarray}
where $\lambda_{t}$ is the tidal deformability parameter and can be defined based on $k_2$, the tidal love number, which is calculated from the system of equations including the TOV one. As is evident, $k_2$ and the tidal deformability  strongly depend on the star's EoS \cite{Hinderer:2007mb,Hinderer:2009ca,Postnikov:2010yn}.
\begin{eqnarray}
\lambda_{t}=\frac23 k_2 R^5
\end{eqnarray}

Unlike $\lambda_{t}$ which has dimension, dimensionless tidal deformability $\Lambda$ can be defined as,
\begin{eqnarray}\label{tidall}
\Lambda=\frac{\lambda_{t}}{M^5}=\frac23 k_2 \left(\frac{R}{M}\right)^5\,.
\end{eqnarray}
where R and M are the radius and mass of the compact star. It should be mentioned that R in a DM admixed NS is the outermost radius of the object which for a DM halo $R=R_{D}$ and for a DM core  $R=R_{B}$.  As an observational constraint on the tidal deformability, we take $\Lambda_{1.4}=190^{+390}_{-120}$ reported by  \cite{Abbott:2018exr} for $M=1.4M\odot$ from the GW170817 event.

\begin{figure}[h]%
\begin{center}
  \parbox{2.4in}{\includegraphics[width=2.4in]{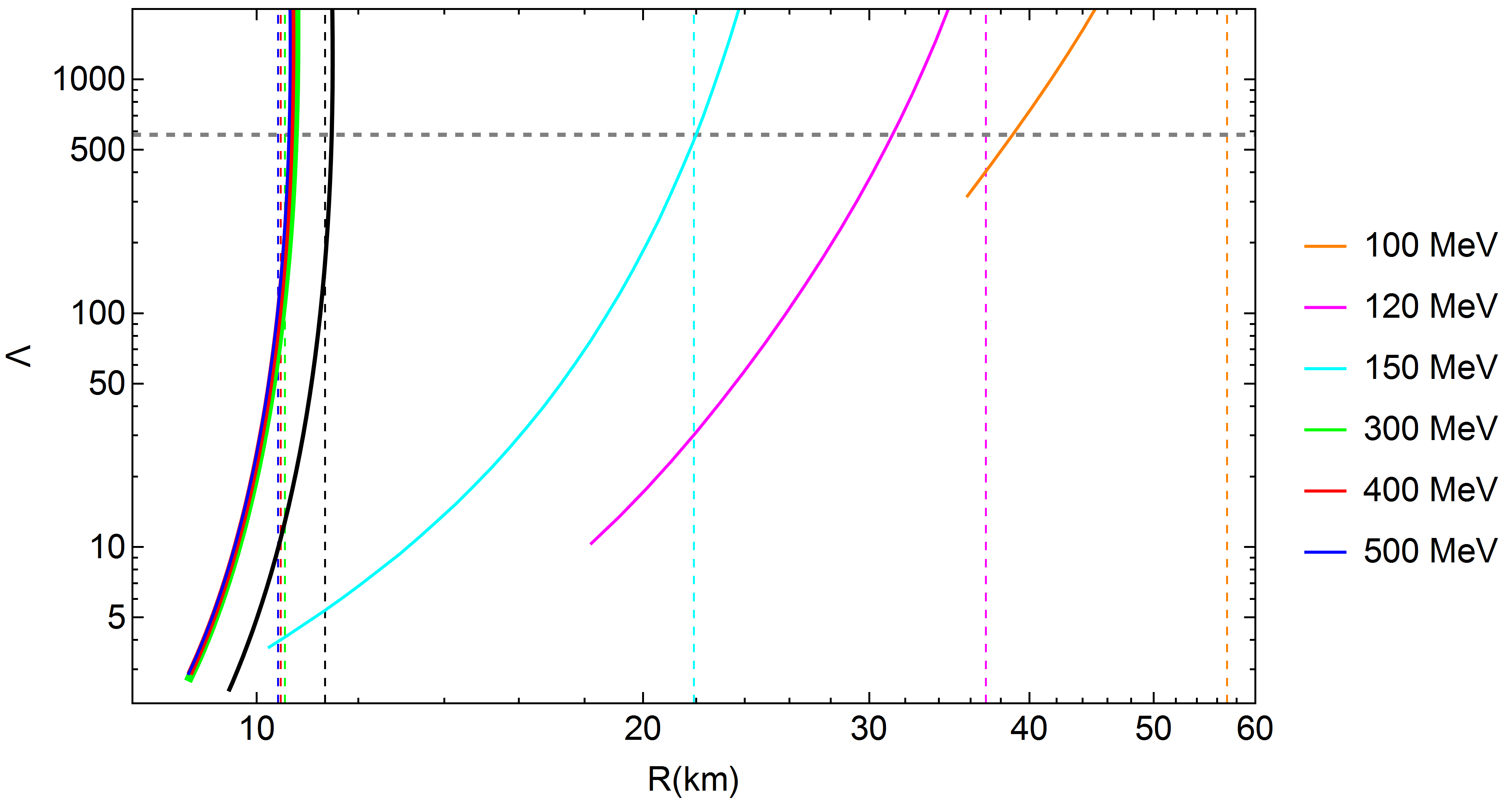}}
  \hspace*{4pt}
  \parbox{2.4in}{\includegraphics[width=2.4in]{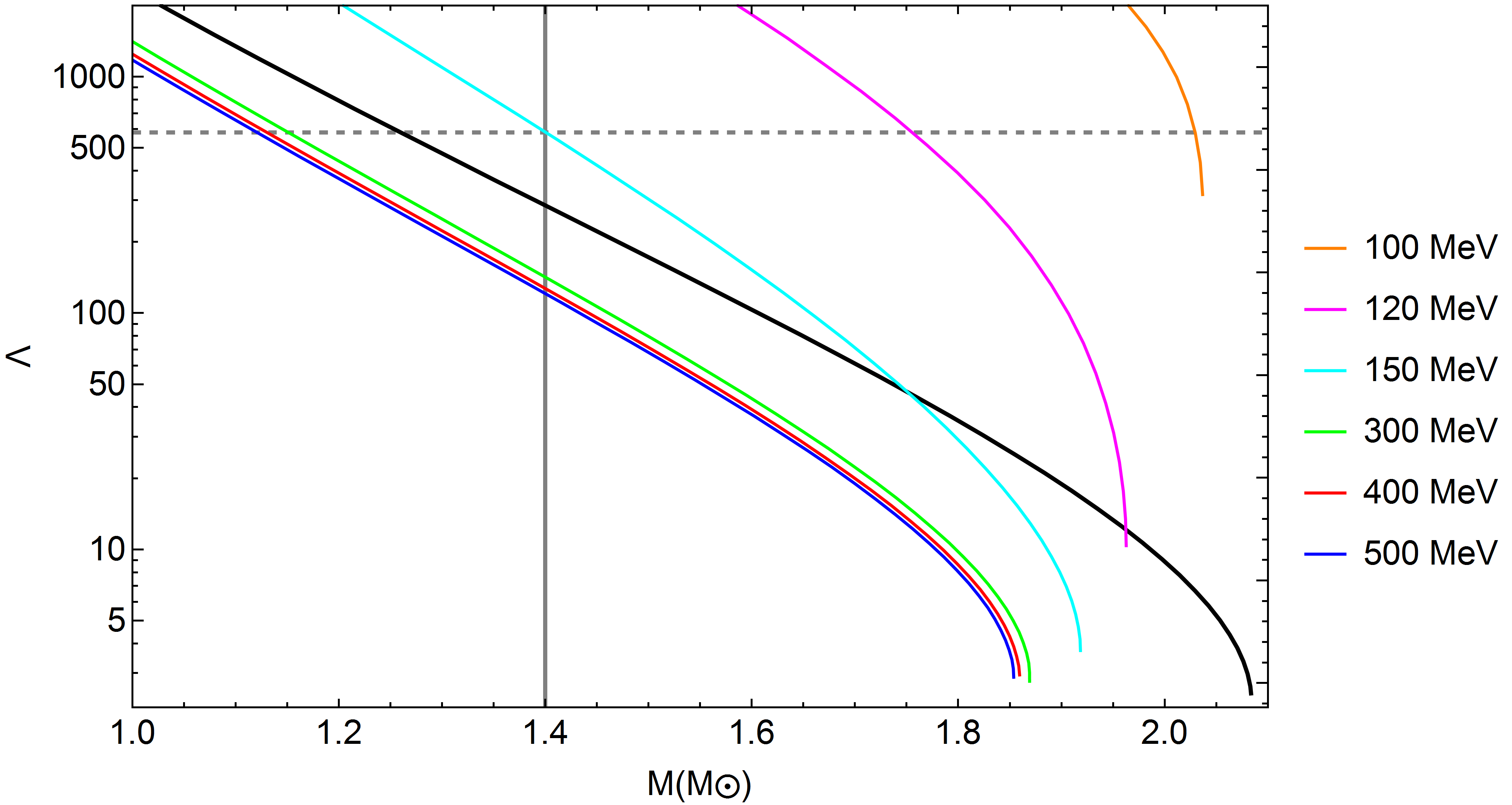}}
  \caption{Tidal deformability ($\Lambda$) as a function of total mass (right) and outermost radius (left) for stable sequences of  DM admixed NSs. Various boson masses are considered, $m_{\chi}=100,120,150$ MeV correspond to a DM halo formation while for $m_{\chi}=300,400,500$ MeV a DM core is formed inside NS. Coupling constant and DM fraction are fixed at $\pi$ and $10\%$, respectively.  }%
  \label{tidal1}
\end{center}
\end{figure}

\begin{figure}[h]%
\begin{center}
  \parbox{2.4in}{\includegraphics[width=2.4in]{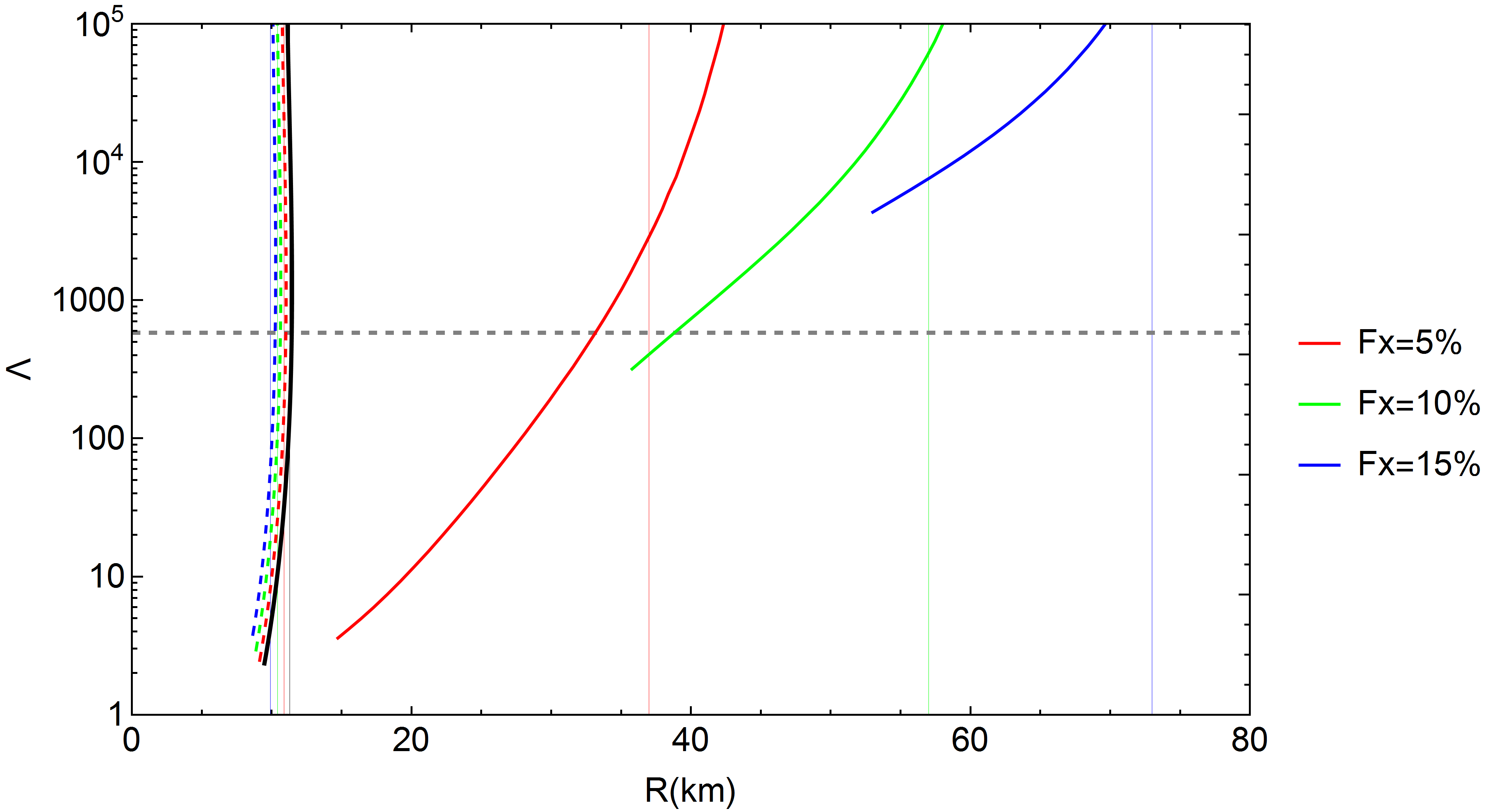}}
  \hspace*{4pt}
  \parbox{2.4in}{\includegraphics[width=2.4in]{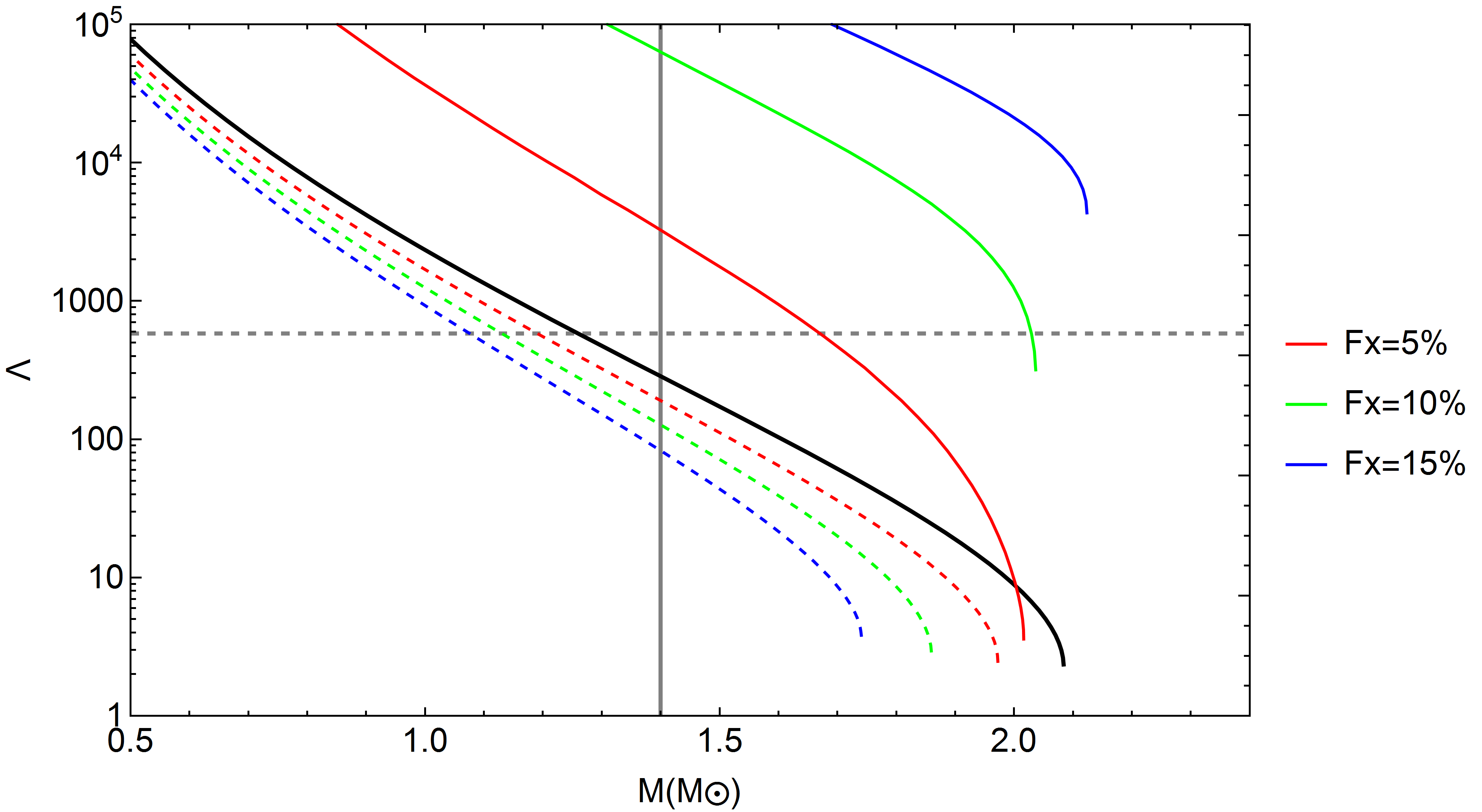}}
  \caption{Tidal deformability ($\Lambda$) as a function of total mass (right) and outermost radius (left) for stable sequences of DM admixed NSs. Two boson masses are considered, $m_{\chi}=100$ MeV (solid lines) and $m_{\chi}=400$ MeV (dashed lines) which correspond to DM halo and DM core formations, respectively. Various DM fractions are considered as labeled.   }%
  \label{tidal2}
\end{center}
\end{figure}

In the following, we investigate the effect of self-interacting bosonic DM, as a DM core or a DM halo, on the tidal deformability $\Lambda$ of a mixed object at various $m_{\chi}$ and $F_{\chi}$. In Figs. \ref{tidal1} and \ref{tidal2} the variation of $\Lambda$ is shown in terms of total mass and radius of DM admixed NSs. In these figures the gray horizontal dashed  lines indicate the LIGO/Virgo upper bound $\Lambda_{1.4}=580$  \cite{Abbott:2018exr}, the gray solid vertical lines show $M_{T}=1.4M_{\odot}$ and the colored dashed and solid vertical lines stand for $R_{1.4}$ radius for the corresponding model parameters. The tidal deformability calculated for the pure baryonic EoS is denoted by the solid black curve and its $\Lambda_{1.4}$ value is about 285 which is well below the LIGO/Virgo constraint.

As a general behaviour in these plots, it can be seen that tidal deformability is a decreasing  function of total mass and  rises by increasing the radius which is related to the definition of this parameter as a function of $R/M$ through Eq. (\ref{tidall}). It follows from the M-R profile of a combined system NS$+$DM, that approaching the maximum mass of the equilibrium sequence decreases the stellar radius and, consequently, $R/M$. In other words the lowest value of $\Lambda$ corresponds to the maximum mass and minimum radius of the DM admixed NS.

 The effect of variation of $\Lambda$ caused by changing the DM mass at fixed coupling constant $\lambda=\pi$ and DM fraction $F_\chi=10\%$ is shown in Fig. \ref{tidal1}. It is seen that for low DM masses $m_{\chi}=100,120,150$ MeV, leading to formation of the DM halo, $\Lambda_{1.4}$ is higher than in the cases of higher $m_{\chi}$ and purely baryonic NS. Indeed, the corresponding $R_{1.4}$ significantly exceeds the value obtained for purely baryonic NS. For $m_{\chi}=300,400,500$ MeV, however, the situation is different. Formation of the DM core reduces the corresponding tidal polarizability, for which the $\Lambda-R$ curves are very similar to purely baryonic case since the radius is determined by $R_B$ (BM EoS). Regarding Fig. \ref{tidal1}, we can conclude that DM halo yields large $\Lambda_{1.4}$, which even can exceed the observational constraint,  while the DM core lowers $\Lambda$  making it consistent with $\Lambda_{1.4}\leq580$. This is related to the effect which was mentioned in the previous section. Namely, DM halo increases the mass and radius of DM admixed NSs while DM core decreases these quantities. It is worth mentioning that GW observations during the inspiral phase of NS-NS coalescence correspond to lower frequencies detectable by Ad. LIGO.  At this regime typical interstellar separation is $r<150$ km. In order to prevent the technical difficulties caused by the overlap of DM halos we restrict their radii as $R_{D}\le75$ km \cite{Nelson_2019,Ellis:2018bkr}. 

To give more insight, Fig. \ref{tidal2} shows modification of tidal polarizability due to variation of the DM fraction from $5\%$ to $15\%$ calculated at fixed $\lambda=\pi$ and $m_{\chi}=100,400$ MeV, corresponding to DM halo and DM core, respectively. As it is seen, higher $F_\chi$ increases $\Lambda_{1.4}$ and $R_{1.4}$ at $m_{\chi}=100$ MeV and decreases theses parameters at $m_{\chi}=400$ MeV. Remarkably, for $m_{\chi}=100$ MeV (solid lines) tidal polarizability of the $M=1.4M\odot$ star exceeds 580 even at $F_{\chi}=5\%$ since in this case $\Lambda$ is more sensitive to DM fraction than at $m_\chi=400$ MeV (dashed lines). Note that in the latter case, the reduction of tidal deformability and $R_{1.4}$ should be consistent with the lower observational limits  $\Lambda_{1.4}\gtrsim70$ and $R_{1.4}\gtrsim11$ km \cite{Abbott:2018exr,Miller:2019cac,Riley:2019yda,Raaijmakers:2021uju,Miller:2021qha}.

\begin{figure}[h]
\begin{center}
\includegraphics[width=4in]{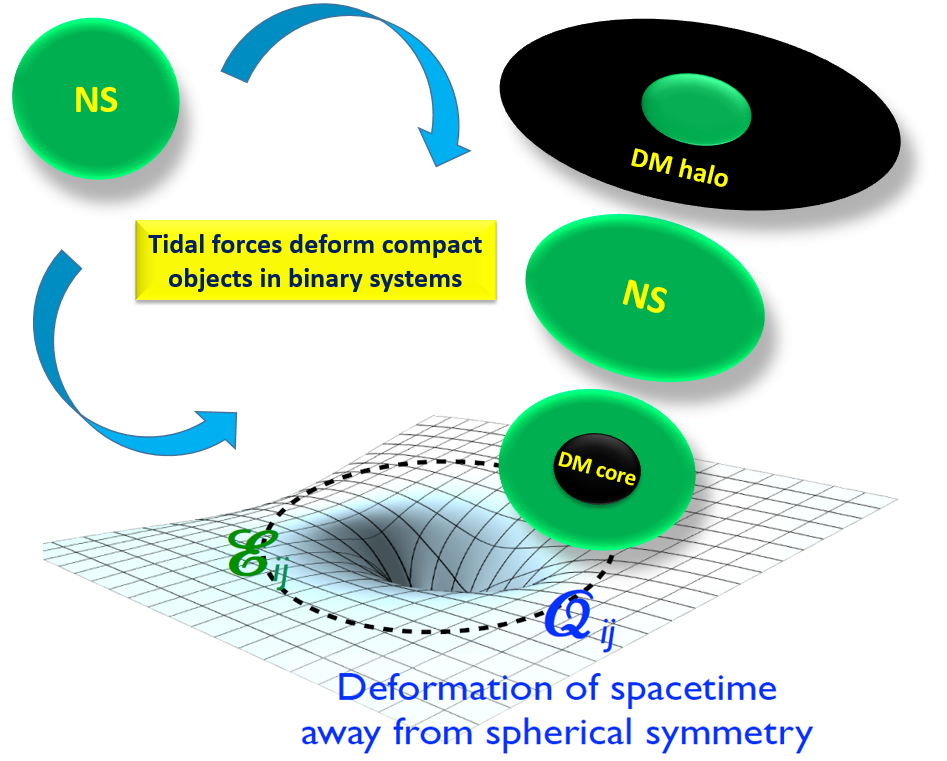}
\end{center}
\caption{The effect of increasing and decreasing of tidal deformability on DM admixed NS's deformation is compared with the pure baryonic NS. It is shown that higher value of $\Lambda$ indicates more deformation in the compact object, therefore a DM halo deforms more in comparison with a pure NS and a DM admixed NS with a DM core.
}
\label{tidal3}
\end{figure}

As the final remark of this section, in Fig. \ref{tidal3}, we show the effect of increasing and decreasing tidal deformability in a NS with DM halo/core and purely BM. It was explained in the beginning of this section that tidal deformability parameter shows how much the compact object is deformed due to the gravitational potential of its companion. Thus in this illustration, we see that DM admixed NS with a DM halo can be more deformed  since it has higher values of $\Lambda$ compared to purely baryonic NS and the DM admixed one with the DM core. In addition, considering tidal love number $k_2$  we note that mixed compact objects with stiffer EoS are more deformable due to the DM halo compared to the ones with softer EoS producing the DM core.

In summary, we note that in full agreement with the previous studies\cite{Nelson_2019,Ellis:2018bkr} DM halo increases tidal deformability, while DM core decreases it. Meanwhile, upper constraint on tidal deformability, $\Lambda_{1.4}=580$, related to GW170817 event \cite{Abbott:2018exr}, has been considered.

\newpage
\section{Conclusion and Outlook}\label{sec5}

Treating DM as a self-repulsing complex scalar field, various properties of single fluid BSs and two-fluid NSs have been studied within the TOV formalism. It is shown that for $\lambda=\pi$, light DM particles ($m_{\chi}\lesssim200$ MeV) form BSs with much larger maximum mass and radius compared with typical NSs, while heavier DM particles lead to formation of BSs with much smaller radius and mass. Furthermore, we showed that  light bosons create a halo around the NS whereas heavy DM particles, for low DM fractions, form a DM core inside the BM component.

The effect of bosonic DM as a  halo/core has been examined by considering the maximum mass, radius and tidal deformability of a DM admixed NS. We have indicated that DM halo formation causes an increase in the aforementioned observable quantities while a DM core reduces all of them. Considering various $m_{\chi}$ and $F_{\chi}$, the maximum mass and tidal deformability of the mixed object have been compared to the latest observational constraints, $M_{max}=2M\odot$ and  $\Lambda_{1.4}\leq580$, inferred from NICER (PSR J0740+6620) and LIGO/Virgo (GW170817) detections.

\begin{figure}[!h]
\begin{center}
\includegraphics[width=4in]{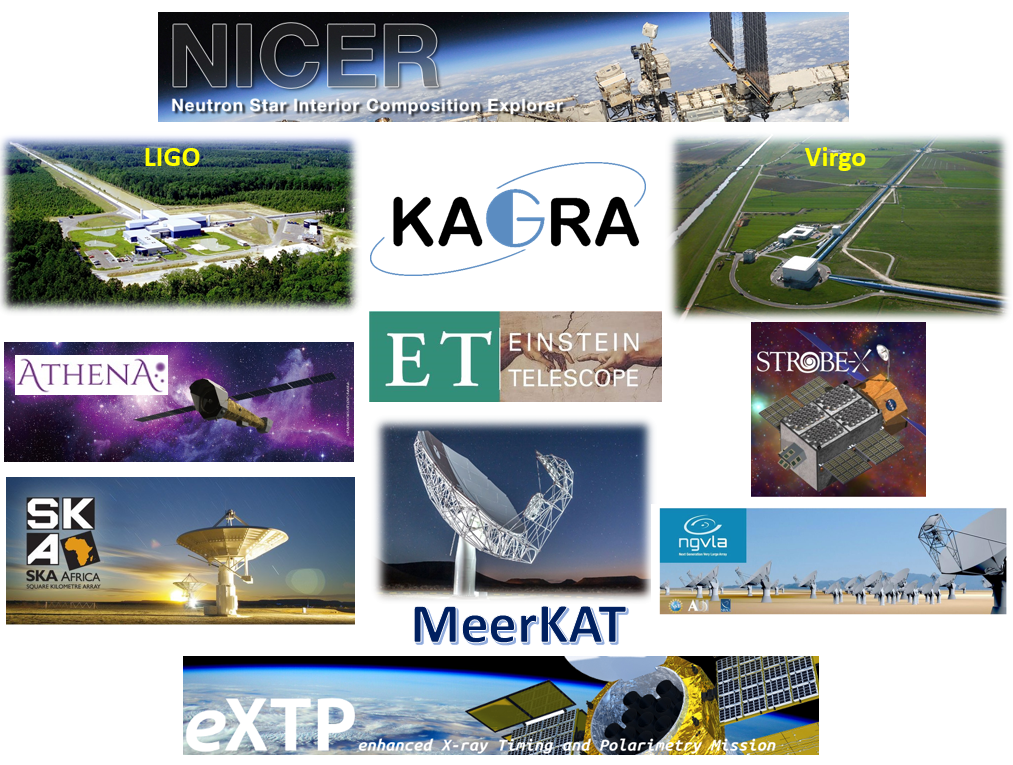}
\end{center}
\caption{Applying various innovative telescopes covering all kinds of observations from GW and X-ray to radio waves provide a unique opportunity for compact objects' research which may solve the puzzle of DM.}
\label{last1}
\end{figure}

Regarding the impact of DM halo and DM core formations on NS's observable parameters and applying the observational limits for NSs' features, one could constrain the parameter space of DM model such as mass and coupling constant and also the amount of DM inside the compact object. In this regard, an extensive investigation has been done recently in \cite{Karkevandi:2021ygv} by the same  authors of the present paper in which a constraint has been imposed on $F_{\chi}$ for sub-GeV DM particles by taking $M_{max}$ and $\Lambda_{1.4}$ bounds. Moreover, as DM core decreases the visible radius of the DM admixed NS ($R_{B}$) and DM halo increases the invisible dark radius of the object ($R_{D}$), radius constraint for typical NSs ($M\approx1.4M\odot$) and the most massive ones ($M\approx2M\odot$) \cite{Miller:2021qha,Riley:2021pdl,Raaijmakers:2021uju} could be utilized to impose more stringent limits on DM parameter space and its fraction. In addition, any unusual observational results of NS properties could be explained by the DM admixed NS model. For instance, there are many effort among  the community to explain the nature of the secondary compact object in  the GW190814\cite{LIGOScientific:2020zkf} event with the mass about $2.6M\odot$ being higher than the maximum NS mass. There are some works explaining the mass of this strange object by the DM core or halo formation within the DM admixed NS scenario \cite{Das:2021yny,Zhang:2020dfi,Lee:2021yyn,DiGiovanni:2021ejn}. Regarding our model, as an example, $m_{\chi}=50$ MeV, $\lambda=\pi$ and $F_{\chi}\approx20\%$ lead to formation of a DM admixed NS with $M_{T}=2.6M\odot$ and detectable radius about 10 km. As a final remark, upcoming modern facilities such as  X-ray (NICER \cite{Watts:2019lbs}, ATHENA \cite{Cassano:2018zwm}, eXTP \cite{eXTP:2018kws} and STROBE-X \cite{STROBE-XScienceWorkingGroup:2019cyd}) and radio (MeerKAT \cite{Bailes:2018azh}, ngVLA \cite{Anu:2013} and SKA\cite{Weltman:2018zrl})  telescopes, as well as GW (LIGO/Virgo/KAGRA \cite{LIGOScientific:2021qlt} and Einstein\cite{Punturo:2010zz,Maggiore:2019uih}) detectors, shown in Fig. \ref{last1},  would provide vast numbers of promising results for NSs' features bringing us to  a golden age of NS investigations and consequently could help us to shed light on the nature of DM and its possible existence in compact objects.

\section*{Acknowledgments}
The article is prepared for the proceedings of the sixteenth Marcel Grossmann meeting (MG16). The authors would like to thank the organizers and conveners of the MG16 for giving the opportunity to present this work and useful discussions during the meeting.  S. S and D.R. K are also grateful to R. Ruffini  whose support made this participation possible. V.S. acknowledges the support from the Funda\c c\~ao para a Ci\^encia e  a Tecnologia (FCT) within the projects No. UID/FIS/04564/2019, No. UID/04564/2020. The work of O.I. was supported by the Polish National Science Center under the grant No. 2019/33/B/ST9/03059.

\bibliographystyle{ws-procs961x669}
\bibliography{ws-pro-sample}

\begin{thebibliography}{100}

\bibitem{Panotopoulos_2017}
G.~Panotopoulos and I.~Lopes, Dark matter effect on realistic equation of state
  in neutron stars, {\em Physical Review D} {\bf 96} (Oct 2017).

\bibitem{Nelson_2019}
A.~E. Nelson, S.~Reddy and D.~Zhou, Dark halos around neutron stars and
  gravitational waves, {\em Journal of Cosmology and Astroparticle Physics}
  {\bf 2019}, p. 012–012 (Jul 2019).

\bibitem{Ellis:2018bkr}
J.~Ellis, G.~Hütsi, K.~Kannike, L.~Marzola, M.~Raidal and V.~Vaskonen, {Dark
  Matter Effects On Neutron Star Properties}, {\em Phys. Rev. D} {\bf 97}, p.
  123007  (2018).

\bibitem{PhysRevD.102.063028}
O.~Ivanytskyi, V.~Sagun and I.~Lopes, Neutron stars: New constraints on
  asymmetric dark matter, {\em Phys. Rev. D} {\bf 102}, p. 063028 (Sep 2020).

\bibitem{Rezaei:2016zje}
Z.~Rezaei, {Study of Dark-Matter Admixed Neutron Stars using the Equation of
  State from the Rotational Curves of Galaxies}, {\em Astrophys. J.} {\bf 835},
  p.~33  (2017).

\bibitem{deLavallaz:2010wp}
A.~de~Lavallaz and M.~Fairbairn, {Neutron Stars as Dark Matter Probes}, {\em
  Phys. Rev. D} {\bf 81}, p. 123521  (2010).

\bibitem{Gresham:2018rqo}
M.~I. Gresham and K.~M. Zurek, {Asymmetric Dark Stars and Neutron Star
  Stability}, {\em Phys. Rev. D} {\bf 99}, p. 083008  (2019).

\bibitem{2021arXiv211106197D}
Y.~{Dengler}, J.~{Schaffner-Bielich} and L.~{Tolos}, {The Second Love Number of
  Dark Compact Planets and Neutron Stars with Dark Matter}, {\em arXiv
  e-prints} , p. arXiv:2111.06197 (November 2021).

\bibitem{Jimenez:2021nmr}
J.~C. Jim\'enez and E.~S. Fraga, {Radial oscillations of quark stars admixed
  with dark matter} (10 2021).

\bibitem{Navarro:1995iw}
J.~F. Navarro, C.~S. Frenk and S.~D.~M. White, {The Structure of cold dark
  matter halos}, {\em Astrophys. J.} {\bf 462}, 563  (1996).

\bibitem{Ruffini_2015}
R.~Ruffini, C.~R. Argüelles and J.~A. Rueda, On the core-halo distribution of
  dark matter in galaxies, {\em Monthly Notices of the Royal Astronomical
  Society} {\bf 451}, p. 622–628 (May 2015).

\bibitem{Arg_elles_2018}
C.~Argüelles, A.~Krut, J.~Rueda and R.~Ruffini, Novel constraints on fermionic
  dark matter from galactic observables i: The milky way, {\em Physics of the
  Dark Universe} {\bf 21}, p. 82–89 (Sep 2018).

\bibitem{2020Univ....6..222D}
A.~{Del Popolo}, M.~{Le Delliou} and M.~{Deliyergiyev}, {Neutron Stars and Dark
  Matter}, {\em Universe} {\bf 6}, p. 222 (November 2020).

\bibitem{Ciancarella:2020msu}
R.~Ciancarella, F.~Pannarale, A.~Addazi and A.~Marciano, {Constraining mirror
  dark matter inside neutron stars}, {\em Phys. Dark Univ.} {\bf 32}, p. 100796
   (2021).

\bibitem{Goldman_2013}
I.~Goldman, R.~Mohapatra, S.~Nussinov, D.~Rosenbaum and V.~Teplitz, Possible
  implications of asymmetric fermionic dark matter for neutron stars, {\em
  Physics Letters B} {\bf 725}, p. 200–207 (Oct 2013).

\bibitem{Ciarcelluti_2011}
P.~Ciarcelluti and F.~Sandin, Have neutron stars a dark matter core?, {\em
  Physics Letters B} {\bf 695}, p. 19–21 (Jan 2011).

\bibitem{Sandin_2009}
F.~Sandin and P.~Ciarcelluti, Effects of mirror dark matter on neutron stars,
  {\em Astroparticle Physics} {\bf 32}, p. 278–284 (Dec 2009).

\bibitem{Kouvaris:2015rea}
C.~Kouvaris and N.~G. Nielsen, {Asymmetric Dark Matter Stars}, {\em Phys. Rev.
  D} {\bf 92}, p. 063526  (2015).

\bibitem{DelPopolo:2020pzh}
A.~Del~Popolo, M.~Deliyergiyev and M.~Le~Delliou, {Solution to the hyperon
  puzzle using dark matter}, {\em Phys. Dark Univ.} {\bf 30}, p. 100622
  (2020).

\bibitem{Deliyergiyev_2019}
M.~Deliyergiyev, A.~Del~Popolo, L.~Tolos, M.~Le~Delliou, X.~Lee and F.~Burgio,
  Dark compact objects: An extensive overview, {\em Physical Review D} {\bf 99}
  (Mar 2019).

\bibitem{Li:2012ii}
A.~Li, F.~Huang and R.-X. Xu, {Too massive neutron stars: The role of dark
  matter?}, {\em Astropart. Phys.} {\bf 37}, 70  (2012).

\bibitem{DelPopolo:2019nng}
A.~Del~Popolo, M.~Deliyergiyev, M.~Le~Delliou, L.~Tolos and F.~Burgio, {On the
  change of old neutron star masses with galactocentric distance}, {\em Phys.
  Dark Univ.} {\bf 28}, p. 100484  (2020).

\bibitem{Kouvaris:2007ay}
C.~Kouvaris, {WIMP Annihilation and Cooling of Neutron Stars}, {\em Phys. Rev.
  D} {\bf 77}, p. 023006  (2008).

\bibitem{Bhat:2019tnz}
S.~A. Bhat and A.~Paul, {Cooling of Dark-Matter Admixed Neutron Stars with
  density-dependent Equation of State}, {\em Eur. Phys. J. C} {\bf 80}, p. 544
  (2020).

\bibitem{Fuller:2014rza}
J.~Fuller and C.~Ott, {Dark Matter-induced Collapse of Neutron Stars: A
  Possible Link Between Fast Radio Bursts and the Missing Pulsar Problem}, {\em
  Mon. Not. Roy. Astron. Soc.} {\bf 450}, L71  (2015).

\bibitem{Acevedo:2020avd}
J.~F. Acevedo, J.~Bramante and A.~Goodman, {Nuclear fusion inside dark matter},
  {\em Phys. Rev. D} {\bf 103}, p. 123022  (2021).

\bibitem{Sedrakian:2015krq}
A.~Sedrakian, {Axion cooling of neutron stars}, {\em Phys. Rev. D} {\bf 93}, p.
  065044  (2016).

\bibitem{Sedrakian:2018kdm}
A.~Sedrakian, {Axion cooling of neutron stars. II. Beyond hadronic axions},
  {\em Phys. Rev. D} {\bf 99}, p. 043011  (2019).

\bibitem{Kaplan:2009ag}
D.~E. Kaplan, M.~A. Luty and K.~M. Zurek, {Asymmetric Dark Matter}, {\em Phys.
  Rev. D} {\bf 79}, p. 115016  (2009).

\bibitem{Shelton:2010ta}
J.~Shelton and K.~M. Zurek, {Darkogenesis: A baryon asymmetry from the dark
  matter sector}, {\em Phys. Rev. D} {\bf 82}, p. 123512  (2010).

\bibitem{Petraki:2013wwa}
K.~Petraki and R.~R. Volkas, {Review of asymmetric dark matter}, {\em Int. J.
  Mod. Phys. A} {\bf 28}, p. 1330028  (2013).

\bibitem{Leung:2011zz}
S.~Leung, M.~Chu and L.~Lin, {Dark-matter admixed neutron stars}, {\em Phys.
  Rev. D} {\bf 84}, p. 107301  (2011).

\bibitem{PhysRevC.89.025803}
Q.-F. Xiang, W.-Z. Jiang, D.-R. Zhang and R.-Y. Yang, Effects of fermionic dark
  matter on properties of neutron stars, {\em Phys. Rev. C} {\bf 89}, p. 025803
  (Feb 2014).

\bibitem{Tolman:1939jz}
R.~C. Tolman, {Static solutions of Einstein's field equations for spheres of
  fluid}, {\em Phys. Rev.} {\bf 55}, 364  (1939).

\bibitem{Oppenheimer:1939ne}
J.~R. Oppenheimer and G.~M. Volkoff, {On Massive neutron cores}, {\em Phys.
  Rev.} {\bf 55}, 374  (1939).

\bibitem{Das:2018frc}
A.~Das, T.~Malik and A.~C. Nayak, {Confronting nuclear equation of state in the
  presence of dark matter using GW170817 observation in relativistic mean field
  theory approach}, {\em Phys. Rev. D} {\bf 99}, p. 043016  (2019).

\bibitem{Das:2020vng}
H.~C. Das, A.~Kumar, B.~Kumar, S.~Kumar~Biswal, T.~Nakatsukasa, A.~Li and S.~K.
  Patra, {Effects of dark matter on the nuclear and neutron star matter}, {\em
  Mon. Not. Roy. Astron. Soc.} {\bf 495}, 4893  (2020).

\bibitem{Sen:2021wev}
D.~Sen and A.~Guha, {Implications of Feebly Interacting Dark Sector on Neutron
  Star Properties and Constraints from GW170817}, {\em Mon. Not. Roy. Astron.
  Soc.} {\bf 504}, p.~3  (2021).

\bibitem{Das:2021wku}
H.~C. Das, A.~Kumar and S.~K. Patra, {Effects of dark matter on the inspiral
  properties of the binary neutron star} (4 2021).

\bibitem{Mukhopadhyay:2016dsg}
S.~Mukhopadhyay, D.~Atta, K.~Imam, D.~Basu and C.~Samanta, {Compact bifluid
  hybrid stars: Hadronic Matter mixed with self-interacting fermionic
  Asymmetric Dark Matter}, {\em Eur. Phys. J. C} {\bf 77}, p. 440  (2017),
  [Erratum: Eur.Phys.J.C 77, 553 (2017)].

\bibitem{Li_2012}
X.~Li, F.~Wang and K.~Cheng, Gravitational effects of condensate dark matter on
  compact stellar objects, {\em Journal of Cosmology and Astroparticle Physics}
  {\bf 2012}, p. 031–031 (Oct 2012).

\bibitem{Li1:2012sg}
X.~Li, T.~Harko and K.~Cheng, {Condensate dark matter stars}, {\em JCAP} {\bf
  06}, p. 001  (2012).

\bibitem{Tolos_2015}
L.~Tolos and J.~Schaffner-Bielich, Dark compact planets, {\em Physical Review
  D} {\bf 92} (Dec 2015).

\bibitem{Kaup:1968zz}
D.~J. Kaup, {Klein-Gordon Geon}, {\em Phys. Rev.} {\bf 172}, 1331  (1968).

\bibitem{Ruffini:1969qy}
R.~Ruffini and S.~Bonazzola, {Systems of selfgravitating particles in general
  relativity and the concept of an equation of state}, {\em Phys. Rev.} {\bf
  187}, 1767  (1969).

\bibitem{Colpi:1986ye}
M.~Colpi, S.~Shapiro and I.~Wasserman, {Boson Stars: Gravitational Equilibria
  of Selfinteracting Scalar Fields}, {\em Phys. Rev. Lett.} {\bf 57}, 2485
  (1986).

\bibitem{Schunck:2003kk}
F.~E. Schunck and E.~W. Mielke, {General relativistic boson stars}, {\em Class.
  Quant. Grav.} {\bf 20}, R301  (2003).

\bibitem{Liebling:2012fv}
S.~L. Liebling and C.~Palenzuela, {Dynamical Boson Stars}, {\em Living Rev.
  Rel.} {\bf 20}, p.~5  (2017).

\bibitem{Visinelli:2021uve}
L.~Visinelli, {Boson Stars and Oscillatons: A Review} (9 2021).

\bibitem{2018NuPhA.970..133B}
K.~A. {Bugaev}, V.~V. {Sagun}, A.~I. {Ivanytskyi}, I.~P. {Yakimenko}, E.~G.
  {Nikonov}, A.~V. {Taranenko} and G.~M. {Zinovjev}, {Going beyond the second
  virial coefficient in the hadron resonance gas model}, {\em Nucl. Phys. A}
  {\bf 970}, 133 (February 2018).

\bibitem{Sagun:2018cpi}
V.~V. Sagun, I.~Lopes and A.~I. Ivanytskyi, {The induced surface tension
  contribution for the equation of state of neutron stars}, {\em Astrophys. J.}
  {\bf 871}, p. 157  (2019).

\bibitem{Sagun11:2018sps}
V.~Sagun, I.~Lopes and A.~Ivanytskyi, {Neutron stars meet constraints from high
  and low energy nuclear physics}, {\em Nucl. Phys. A} {\bf 982}, 883  (2019).

\bibitem{Quddus:2019ghy}
A.~Quddus, G.~Panotopoulos, B.~Kumar, S.~Ahmad and S.~K. Patra, {GW170817
  constraints on the properties of a neutron star in the presence of WIMP dark
  matter}, {\em J. Phys. G} {\bf 47}, p. 095202  (2020).

\bibitem{Zhang:2020dfi}
K.~Zhang and F.-L. Lin, {Constraint on hybrid stars with gravitational wave
  events}, {\em Universe} {\bf 6}, p. 231  (2020).

\bibitem{LeTiec:2020spy}
A.~Le~Tiec and M.~Casals, {Spinning Black Holes Fall in Love}, {\em Phys. Rev.
  Lett.} {\bf 126}, p. 131102  (2021).

\bibitem{Das_2019}
A.~Das, T.~Malik and A.~C. Nayak, Confronting nuclear equation of state in the
  presence of dark matter using gw170817 observation in relativistic mean field
  theory approach, {\em Physical Review D} {\bf 99} (Feb 2019).

\bibitem{Ellis:2017jgp}
J.~Ellis, A.~Hektor, G.~Hütsi, K.~Kannike, L.~Marzola, M.~Raidal and
  V.~Vaskonen, {Search for Dark Matter Effects on Gravitational Signals from
  Neutron Star Mergers}, {\em Phys. Lett. B} {\bf 781}, 607  (2018).

\bibitem{Bezares:2019jcb}
M.~Bezares, D.~Vigan\`o and C.~Palenzuela, {Gravitational wave signatures of
  dark matter cores in binary neutron star mergers by using numerical
  simulations}, {\em Phys. Rev. D} {\bf 100}, p. 044049  (2019).

\bibitem{Bezares:2018qwa}
M.~Bezares and C.~Palenzuela, {Gravitational Waves from Dark Boson Star binary
  mergers}, {\em Class. Quant. Grav.} {\bf 35}, p. 234002  (2018).

\bibitem{Horowitz:2019aim}
C.~Horowitz and S.~Reddy, {Gravitational Waves from Compact Dark Objects in
  Neutron Stars}, {\em Phys. Rev. Lett.} {\bf 122}, p. 071102  (2019).

\bibitem{Bauswein:2020kor}
A.~Bauswein, G.~Guo, J.-H. Lien, Y.-H. Lin and M.-R. Wu, {Compact Dark Objects
  in Neutron Star Mergers} (12 2020).

\bibitem{LIGOScientific:2017vwq}
B.~P. Abbott {\em et~al.}, {GW170817: Observation of Gravitational Waves from a
  Binary Neutron Star Inspiral}, {\em Phys. Rev. Lett.} {\bf 119}, p. 161101
  (2017).

\bibitem{LIGOScientific:2020aai}
B.~P. Abbott {\em et~al.}, {GW190425: Observation of a Compact Binary
  Coalescence with Total Mass $\sim 3.4 M_{\odot}$}, {\em Astrophys. J. Lett.}
  {\bf 892}, p.~L3  (2020).

\bibitem{Riley:2021pdl}
T.~E. Riley {\em et~al.}, {A NICER View of the Massive Pulsar PSR J0740+6620
  Informed by Radio Timing and XMM-Newton Spectroscopy}, {\em Astrophys. J.
  Lett.} {\bf 918}, p. L27  (2021).

\bibitem{Abbott_2017}
B.~Abbott, R.~Abbott, T.~Abbott, F.~Acernese, K.~Ackley, C.~Adams, T.~Adams,
  P.~Addesso, R.~Adhikari, V.~Adya and et~al., Gw170817: Observation of
  gravitational waves from a binary neutron star inspiral, {\em Physical Review
  Letters} {\bf 119} (Oct 2017).

\bibitem{Abbott:2018exr}
B.~P. Abbott {\em et~al.}, {GW170817: Measurements of neutron star radii and
  equation of state}, {\em Phys. Rev. Lett.} {\bf 121}, p. 161101  (2018).

\bibitem{Maselli:2017vfi}
A.~Maselli, P.~Pnigouras, N.~G. Nielsen, C.~Kouvaris and K.~D. Kokkotas, {Dark
  stars: gravitational and electromagnetic observables}, {\em Phys. Rev. D}
  {\bf 96}, p. 023005  (2017).

\bibitem{PhysRevD.68.023511}
A.~Arbey, J.~Lesgourgues and P.~Salati, Galactic halos of fluid dark matter,
  {\em Phys. Rev. D} {\bf 68}, p. 023511 (Jul 2003).

\bibitem{Suarez:2013iw}
A.~Su\'arez, V.~H. Robles and T.~Matos, {A Review on the Scalar
  Field/Bose-Einstein Condensate Dark Matter Model}, {\em Astrophys. Space Sci.
  Proc.} {\bf 38}, 107  (2014).

\bibitem{Mielke:2000mh}
E.~W. Mielke and F.~E. Schunck, {Boson stars: Alternatives to primordial black
  holes?}, {\em Nucl. Phys. B} {\bf 564}, 185  (2000).

\bibitem{Chavanis:2011cz}
P.-H. Chavanis and T.~Harko, {Bose-Einstein Condensate general relativistic
  stars}, {\em Phys. Rev. D} {\bf 86}, p. 064011  (2012).

\bibitem{Karkevandi:2021ygv}
D.~{Rafiei Karkevandi}, S.~{Shakeri}, V.~{Sagun} and O.~{Ivanytskyi}, {Bosonic
  Dark Matter in Neutron Stars and its Effect on Gravitational Wave Signal},
  {\em arXiv e-prints} , p. arXiv:2109.03801 (September 2021).

\bibitem{Pacilio:2020jza}
C.~Pacilio, M.~Vaglio, A.~Maselli and P.~Pani, {Gravitational-wave detectors as
  particle-physics laboratories: Constraining scalar interactions with a
  coherent inspiral model of boson-star binaries}, {\em Phys. Rev. D} {\bf
  102}, p. 083002  (2020).

\bibitem{Li:2012sg}
X.~Li, T.~Harko and K.~Cheng, {Condensate dark matter stars}, {\em JCAP} {\bf
  06}, p. 001  (2012).

\bibitem{Chavanis:2019bnu}
P.-H. Chavanis, {Mass-radius relation of self-gravitating Bose-Einstein
  condensates with a central black hole}, {\em Eur. Phys. J. Plus} {\bf 134},
  p. 352  (2019).

\bibitem{Dalfovo_1999}
F.~Dalfovo, S.~Giorgini, L.~P. Pitaevskii and S.~Stringari, Theory of
  bose-einstein condensation in trapped gases, {\em Reviews of Modern Physics}
  {\bf 71}, p. 463–512 (Apr 1999).

\bibitem{Rogel_Salazar_2013}
J.~Rogel-Salazar, The gross–pitaevskii equation and bose–einstein
  condensates, {\em European Journal of Physics} {\bf 34}, p. 247–257 (Jan
  2013).

\bibitem{pethick_smith_2008}
C.~J. Pethick and H.~Smith, {\em Bose-Einstein Condensation in Dilute Gases}, 2
  edn. (Cambridge University Press, 2008).

\bibitem{Antoniadis:2013pzd}
J.~Antoniadis {\em et~al.}, {A Massive Pulsar in a Compact Relativistic
  Binary}, {\em Science} {\bf 340}, p. 6131  (2013).

\bibitem{Cromartie:2019kug}
H.~T. Cromartie {\em et~al.}, {Relativistic Shapiro delay measurements of an
  extremely massive millisecond pulsar}, {\em Nature Astron.} {\bf 4}, 72
  (2019).

\bibitem{AmaroSeoane:2010qx}
P.~Amaro-Seoane, J.~Barranco, A.~Bernal and L.~Rezzolla, {Constraining scalar
  fields with stellar kinematics and collisional dark matter}, {\em JCAP} {\bf
  11}, p. 002  (2010).

\bibitem{2014NuPhA.924...24S}
V.~V. Sagun, A.~I. Ivanytskyi, K.~A. Bugaev and I.~N. Mishustin, {The
  statistical multifragmentation model for liquid-gas phase transition with a
  compressible nuclear liquid}, {\em Nucl. Phys. A} {\bf 924}, 24  (2014).

\bibitem{Ivanytskyi:2017pkt}
A.~I. Ivanytskyi, K.~A. Bugaev, V.~V. Sagun, L.~V. Bravina and E.~E. Zabrodin,
  {Influence of flow constraints on the properties of the critical endpoint of
  symmetric nuclear matter}, {\em Phys. Rev. C} {\bf 97}, p. 064905  (2018).

\bibitem{2018EPJA...54..100S}
V.~V. {Sagun}, K.~A. {Bugaev}, A.~I. {Ivanytskyi}, I.~P. {Yakimenko}, E.~G.
  {Nikonov}, A.~V. {Taranenko}, C.~{Greiner}, D.~B. {Blaschke} and G.~M.
  {Zinovjev}, {Hadron resonance gas model with induced surface tension}, {\em
  European Physical Journal A} {\bf 54}, p. 100 (June 2018).

\bibitem{Sagun:2020qvc}
V.~Sagun, G.~Panotopoulos and I.~Lopes, {Asteroseismology: radial oscillations
  of neutron stars with realistic equation of state}, {\em Phys. Rev. D} {\bf
  101}, p. 063025  (2020).

\bibitem{Raithel:2018ncd}
C.~Raithel, F.~\"Ozel and D.~Psaltis, {Tidal deformability from GW170817 as a
  direct probe of the neutron star radius}, {\em Astrophys. J. Lett.} {\bf
  857}, p. L23  (2018).

\bibitem{Chatziioannou:2020pqz}
K.~Chatziioannou, {Neutron star tidal deformability and equation of state
  constraints}, {\em Gen. Rel. Grav.} {\bf 52}, p. 109  (2020).

\bibitem{Most:2018hfd}
E.~R. Most, L.~R. Weih, L.~Rezzolla and J.~Schaffner-Bielich, {New constraints
  on radii and tidal deformabilities of neutron stars from GW170817}, {\em
  Phys. Rev. Lett.} {\bf 120}, p. 261103  (2018).

\bibitem{De:2018uhw}
S.~De, D.~Finstad, J.~M. Lattimer, D.~A. Brown, E.~Berger and C.~M. Biwer,
  {Tidal Deformabilities and Radii of Neutron Stars from the Observation of
  GW170817}, {\em Phys. Rev. Lett.} {\bf 121}, p. 091102  (2018), [Erratum:
  Phys.Rev.Lett. 121, 259902 (2018)].

\bibitem{Hinderer:2007mb}
T.~Hinderer, {Tidal Love numbers of neutron stars}, {\em Astrophys. J.} {\bf
  677}, 1216  (2008).

\bibitem{Hinderer:2009ca}
T.~Hinderer, B.~D. Lackey, R.~N. Lang and J.~S. Read, {Tidal deformability of
  neutron stars with realistic equations of state and their gravitational wave
  signatures in binary inspiral}, {\em Phys. Rev. D} {\bf 81}, p. 123016
  (2010).

\bibitem{Postnikov:2010yn}
S.~Postnikov, M.~Prakash and J.~M. Lattimer, {Tidal Love Numbers of Neutron and
  Self-Bound Quark Stars}, {\em Phys. Rev. D} {\bf 82}, p. 024016  (2010).

\bibitem{Zhao:2018nyf}
T.~Zhao and J.~M. Lattimer, {Tidal Deformabilities and Neutron Star Mergers},
  {\em Phys. Rev. D} {\bf 98}, p. 063020  (2018).

\bibitem{Han:2018mtj}
S.~Han and A.~W. Steiner, {Tidal deformability with sharp phase transitions in
  (binary) neutron stars}, {\em Phys. Rev. D} {\bf 99}, p. 083014  (2019).

\bibitem{Lackey:2014fwa}
B.~D. Lackey and L.~Wade, {Reconstructing the neutron-star equation of state
  with gravitational-wave detectors from a realistic population of inspiralling
  binary neutron stars}, {\em Phys. Rev. D} {\bf 91}, p. 043002  (2015).

\bibitem{Miller:2019cac}
M.~C. Miller {\em et~al.}, {PSR J0030+0451 Mass and Radius from $NICER$ Data
  and Implications for the Properties of Neutron Star Matter}, {\em Astrophys.
  J. Lett.} {\bf 887}, p. L24  (2019).

\bibitem{Riley:2019yda}
T.~E. Riley {\em et~al.}, {A $NICER$ View of PSR J0030+0451: Millisecond Pulsar
  Parameter Estimation}, {\em Astrophys. J. Lett.} {\bf 887}, p. L21  (2019).

\bibitem{Raaijmakers:2021uju}
G.~Raaijmakers, S.~K. Greif, K.~Hebeler, T.~Hinderer, S.~Nissanke, A.~Schwenk,
  T.~E. Riley, A.~L. Watts, J.~M. Lattimer and W.~C.~G. Ho, {Constraints on the
  dense matter equation of state and neutron star properties from NICER's
  mass-radius estimate of PSR J0740+6620 and multimessenger observations} (5
  2021).

\bibitem{Miller:2021qha}
M.~C. Miller {\em et~al.}, {The Radius of PSR J0740+6620 from NICER and
  XMM-Newton Data}, {\em Astrophys. J. Lett.} {\bf 918}, p. L28  (2021).

\bibitem{LIGOScientific:2020zkf}
R.~Abbott {\em et~al.}, {GW190814: Gravitational Waves from the Coalescence of
  a 23 Solar Mass Black Hole with a 2.6 Solar Mass Compact Object}, {\em
  Astrophys. J. Lett.} {\bf 896}, p. L44  (2020).

\bibitem{Das:2021yny}
H.~C. Das, A.~Kumar and S.~K. Patra, {Dark matter admixed neutron star as a
  possible compact component in the GW190814 merger event}, {\em Phys. Rev. D}
  {\bf 104}, p. 063028  (2021).

\bibitem{Lee:2021yyn}
B.~K.~K. Lee, M.-c. Chu and L.-M. Lin, {Can the GW190814 secondary component be
  a bosonic dark matter admixed compact star?} (10 2021).

\bibitem{DiGiovanni:2021ejn}
F.~Di~Giovanni, N.~Sanchis-Gual, P.~Cerd\'a-Dur\'an and J.~A. Font, {Can
  fermion-boson stars reconcile multi-messenger observations of compact stars?}
  (10 2021).

\bibitem{Watts:2019lbs}
A.~L. Watts, {Constraining the neutron star equation of state using Pulse
  Profile Modeling}, {\em AIP Conf. Proc.} {\bf 2127}, p. 020008  (2019).

\bibitem{Cassano:2018zwm}
R.~Cassano {\em et~al.}, {SKA-Athena Synergy White Paper} (7 2018).

\bibitem{eXTP:2018kws}
J.~J.~M. in~'t Zand {\em et~al.}, {Observatory science with eXTP}, {\em Sci.
  China Phys. Mech. Astron.} {\bf 62}, p. 029506  (2019).

\bibitem{STROBE-XScienceWorkingGroup:2019cyd}
P.~S. Ray {\em et~al.}, {STROBE-X: X-ray Timing and Spectroscopy on Dynamical
  Timescales from Microseconds to Years} (3 2019).

\bibitem{Bailes:2018azh}
M.~Bailes {\em et~al.}, {MeerTime - the MeerKAT Key Science Program on Pulsar
  Timing}, {\em PoS} {\bf MeerKAT2016}, p. 011  (2018).

\bibitem{Anu:2013}
R.~J. Selina, The next generation very large array: A technical overview
  \url{https://arxiv.org/ftp/arxiv/papers/1806/1806.08405.pdf},  (2018).

\bibitem{Weltman:2018zrl}
A.~Weltman {\em et~al.}, {Fundamental physics with the Square Kilometre Array},
  {\em Publ. Astron. Soc. Austral.} {\bf 37}, p. e002  (2020).

\bibitem{LIGOScientific:2021qlt}
R.~Abbott {\em et~al.}, {Observation of Gravitational Waves from Two Neutron
  Star\textendash{}Black Hole Coalescences}, {\em Astrophys. J. Lett.} {\bf
  915}, p.~L5  (2021).

\bibitem{Punturo:2010zz}
M.~Punturo {\em et~al.}, {The Einstein Telescope: A third-generation
  gravitational wave observatory}, {\em Class. Quant. Grav.} {\bf 27}, p.
  194002  (2010).

\bibitem{Maggiore:2019uih}
M.~Maggiore {\em et~al.}, {Science Case for the Einstein Telescope}, {\em JCAP}
  {\bf 03}, p. 050  (2020).

\end{thebibliography}

\end{document}